\documentclass[conference,10pt,twocolumns]{IEEEtran}

\usepackage{graphicx}
\usepackage[center]{caption}
\usepackage{epsfig,latexsym,amsmath,amsfonts}
\usepackage{amssymb}
\usepackage{placeins}
\usepackage{verbatim}
\usepackage{cite,url}
\usepackage{epsf,bm}
\usepackage{epstopdf}
\usepackage{breqn}
\usepackage{subcaption}

\begin{document}

\title{A Family of Blind Symbol Rate Estimators using Zero Crossing Detection over Unknown Multipath Fading Channels}

\author{\large  Mahmoud Elgenedy$^*$, Ayman Elezabi$^{\dagger}$ \vspace{0.1in} \\
\normalsize   \begin{tabular}{c}
$^*$University of Texas at Dallas\\
$^{\dagger}$Electronics and Communications Engineering Department, American University in Cairo, Egypt.\\ \small
email: mahmoud.elgenedy@utdallas.edu, aelezabi@aucegypt.edu
\end{tabular}
}
\maketitle

\begin{abstract}
In this paper we introduce a new and simple method to estimate the symbol rate for single carrier systems of unknown modulation. The introduced technique detects the symbol rate from a continuous range of symbol rates, i.e. it does not assume a finite set of candidate symbol rates. The method does not require any knowledge about system parameters and is therefore totally blind. The method belongs to the family of autocorrelation-based symbol rate estimators, yet, unlike many such schemes, its performance is insensitive to the value of the roll-off factor. We then propose a simple method for frequency offset compensation also based on the autocorrelation function. Moreover, we study the effect of multipath channel on the estimator performance and propose different enhancements including exploiting more than one zero crossing. The proposed estimator is implemented as part of a complete DVB-C \cite{ETSIDVBC} receiver and is verified using simulations and results in robust performance even at low SNR and high frequency offsets.


\end{abstract}

\section{Introduction}
Symbol rate estimation (SRE) appears in many problems including commercial systems such as DVB-C, DVB-S and DVB-S2. In the cable standard, the cable channel width is 8 MHz in virtually all European countries. However, no symbol rate was fixed for cable TV. The reason for this was the desire to be able to feed satellite signals to cable networks without any processing at the cable head-end. As a result, variable symbol-rate modem design was the best solution for cable TV systems \cite{Karam}. Also, a demodulator with variable-rate timing recovery finds application in mobile and military communications when it is combined with certain types of multi-rate sources and channel coding techniques \cite{Lee}.
Blind symbol rate estimation has been considered in the literature from different point of views. The use of a filter bank matched to different pulse shapes has been proposed in \cite{Lee} through an ad-hoc approach and in \cite{Meyr} based on the Maximum Likelihood (ML) criterion. Less complicated techniques based on cyclic autocorrelation function are proposed in \cite{Gardner,Dandawate,Mazet,Wanxue}. The basic idea behind the cyclic autocorrelation approach is that the autocorrelation function of a linearly modulated sequence is a periodic function in time with a period equal to the symbol period. The authors in \cite{Gardner} express the autocorrelation function as a Fourier series and extract the symbol period from the coefficients of the Fourier series expansion.
Cyclic correlation based symbol rate estimators are more robust against frequency offset and frequency selective fading channels and simpler than matched filter estimators. However, performance degradation occurs for low SNRs and small roll-off factors and may totally fail when the roll-off factor approaches zero. In \cite{Dandawate}, the authors modified the method of \cite{Gardner} and proposed the use of a weighing matrix in computation of the cyclic autocorrelation function to detect the cycle frequencies for pulse shapes having small roll-offs, but in the absence of carrier frequency offset. The authors in \cite{Mazet} used the concept proposed in \cite{Dandawate} to estimate the symbol rate of a linearly modulated signal. However, performance in the low SNR range is still a problem. In \cite{Wanxue}, the authors propose a modified cyclic autocorrelation approach for the symbol rate estimation of M-PSK signals. Asymptotic analysis of the cyclic correlation based symbol-rate estimator is established in \cite{Ciblat}. In \cite{Mosquera}, the ML approach is proved to exploit both the shaping pulse and cyclic correlation. This technique is more robust at low SNR but still suffers a bias in the estimate, that depends on the roll-off factor, at low SNR. If the roll-off factor is unknown, higher error will result. All of the previous techniques estimate the symbol rate by maximizing certain criteria over all possible rates and hence high complexity is expected in case of large sets of symbol rates. In \cite{Ciblat,Mosquera}, this problem is addressed using a two-step search technique where a coarse search is carried out first on a small set of rates, then a finer search around this coarse estimate is performed. Symbol rate estimation from the autocorrelation function is presented in \cite{Chan}. However, the algorithm requires some a priori knowledge (not exact) about the symbol rate and performance is not given in practical cases as a very large sampling rate is assumed.
In this paper we provide a new and simple technique for estimating the symbol rate blindly from a continuous range of symbol rates. The proposed technique is robust at low SNR and does not require prior information on roll-off factor, pulse shape, or any other system parameters. The method relies on obtaining the time-averaged autocorrelation function of the received sequence. Since the received signal is wide-sense cyclostationary, the cycle average autocorrelation is estimated, and a simple method is presented for detecting the first zero-crossing using interpolation. Hence, the only requirement is that the transmitted pulse shape contains a zero crossing, which is satisfied for virtually all practical pulse shapes.
Building on our initial findings in \cite{6655324}, in this paper, our main contributions are:
\begin{itemize}
\item We present a detailed analysis for the different types of errors to beter understand the behavior and limits of the proposed symbol rate estimator.
\item We propose different enhancements to the current algorithm by exploiting more than one zero crossings.
\item We explore the realistic channel conditions reported in the Nordig standard for cable broadcasting.

\end{itemize}

The rest of the paper is organized as follows. In the next section we present the system model. In section III the new algorithm details are explained. Simulation results are shown in section IV based on the DVB-C system. The frequency offset problem and compensation is presented at section V. Finally, conclusions and current and future work are discussed in section VI.


\section{System Model}\label{Sys}
\subsection{Transmitted Signal}

Let $y(t)$ denotes the complex baseband received signal,
\begin{equation}\label{syseqn}
    y(t)\ =\ e^{-j2\pi f_{o} t}\ \sum_{k\in Z} a_k g(t-kT)+\omega (t)
\end{equation}
where $\{a_k\}_{k\in Z}$ is a zero-mean independent and identically distributed (i.i.d.) sequence of QAM symbols. The pulse shaping filter $g(t)$ is a square-root raised cosine (SRRC) pulse, such that $G(f)=0,\ |f|>(1+\alpha)/2T$, for a roll-off factor $0\leq \alpha \leq 1$ and baud rate $1/T$. $\omega (t)$ represents the narrowband zero mean additive Gaussian noise; i.e. it is a complex Gaussian circular random process. As stated earlier, the method is applicable for other pulse shapes, but for convenience we henceforth consider the square root raised cosine (SRRC) pulse due to its widespread application.

\subsection{Receiver Model}

\begin{figure}
\centering
\includegraphics[width=3.2in]{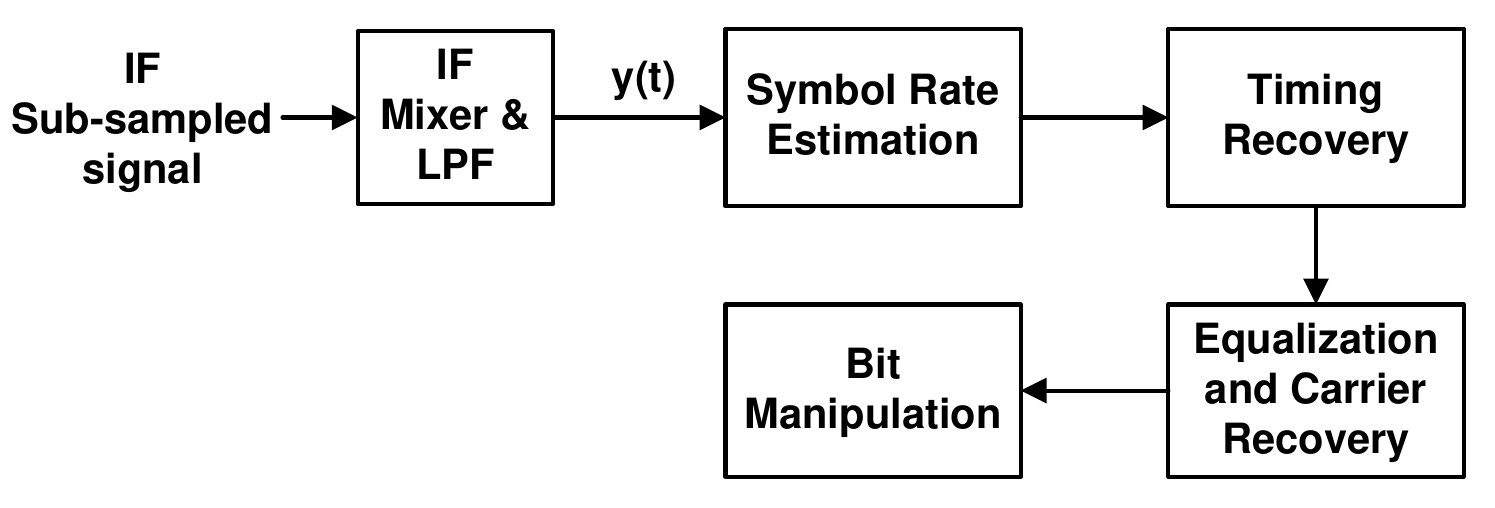}
\caption{Block diagram for the variable rate DVB-C receiver with blind symbol rate estimation}
\label{Fig1}
\end{figure}

Fig. \ref{Fig1} shows the block diagram of the variable rate DVB-C receiver with blind symbol rate estimation. The incoming IF sub-sampled signal is passed on to the IF Mixer followed by a LPF to get the received complex base band signal $y(t)$, where the IF frequency is 36 MHz and the sampling frequency is 56 MHz.
It is worth mentioning that the SRE employed is independent of the remainder of the receiver chain shown in Fig. \ref{Fig1}, which is shown here for completeness. Note that a time tracking block is necessary, even if the symbol rate is known beforehand, to compensate for sample timing inaccuracy as well as residual SRE error in the case of unknown symbol rate. The importance of this notion will become apparent later when we compare practical SRE performance.

\subsection{Multi-path Fading Channel}

To consider all practical cases, we check our algorithm performance in case of multipath fading channel. We focus our work here with the DVB-C channel. As defined in \cite{Nordig}, the DVB-C channel is an echo channel with different gain/delay configurations stated in table I.

The main problem of the fading channel is due to the distortion occurs in the estimated RRC pulse. When the delay between the two path is smaller than the symbol duration, this will highly distort the Sinc pulse as the pulse main lobe will be spreaded. For the DVB-C, this will be the case for the first four echo channels for all supported symbol rates. Also, it will be the case for all echo channels for the symbol rates below 3 MSymbols/sec. Figure \ref{Sinc_Distortion_oneside} shows the distorted Sinc pulse due to different echo channels in the case of single pulse.

\begin{center}

\begin{table}
\caption{DVB-C Echo Channel Paths defined in NorDig Specification \cite{Nordig}}

\begin{center}

\begin{tabular}{|l|l|l|}
\hline
Path No. & Att. [dB] & Delay [ns]\tabularnewline
\hline
\hline
0 & 12 & 0\tabularnewline
\hline
1 & 12.6 & 40\tabularnewline
\hline
2 & 13.7 & 50\tabularnewline
\hline
3 & 19.4 & 100\tabularnewline
\hline
4 & 25 & 150\tabularnewline
\hline
5 & 30.7 & 200\tabularnewline
\hline
6 & 36.3 & 250\tabularnewline
\hline
7 & 39.7 & 280\tabularnewline
\hline
8 & 42 & 300\tabularnewline
\hline
9 & 42 & 350\tabularnewline
\hline
\end{tabular}

\end{center}

\end{table}

\label{table:t1}

\end{center}

\section{SYMBOL RATE ESTIMATION}

The proposed SRE works by first obtaining the time-averaged autocorrelation function of the received sequence. It is well-known that the received signal is wide-sense cyclostationary. We therefore seek the autocorrelation function averaged over one period. As shown below, this results in the composite pulse shape after matched filtering, which in our case is the RC pulse, from which the symbol rate can be estimated.
The autocorrelation function is
\begin{equation}\label{eqautocorr}
    R_y (t,\tau)\ =\ E[y(t)y^{\ast} (t+\tau)]
\end{equation}

Using $y(t)$ from \ref{eqautocorr} without noise or frequency offset, and after some manipulation,

\begin{equation}
    R_y(t,\tau)\ =\ \sum_{n=-\infty}^{\infty} g(t-nT)g(t+\tau-nT)E[{a_n}^2]
\end{equation}

$E[{a_n}^2]$ is constant, and after further manipulations the autocorrelation averaged over a period T, within a constant factor, is

\begin{align}
    \bar{R}_y (\tau)\ &=\ \frac{1}{T}\ \int\limits_{-T/2}^{T/2} \sum_{n=-\infty}^{\infty} \ g(t-nT)g(t+\tau-nT)\ dt \nonumber \\
    &=\ \frac{1}{T} \sum_{n=-\infty}^{\infty}\ \int\limits_{-T/2}^{T/2} g(t-nT)g(t+\tau-nT)\ dt \nonumber \\
    &=\ \frac{1}{T} \int\limits_{-\infty}^{\infty} g(t)g(t+\tau)\ dt
\end{align}

Since $g(t)$ is symmetric,

\begin{equation}
    \bar{R}_y (\tau)\ =\ g(t)*g^{\ast} (t) \nonumber
\end{equation}

\begin{figure}
\centering
\includegraphics[width=3.2in]{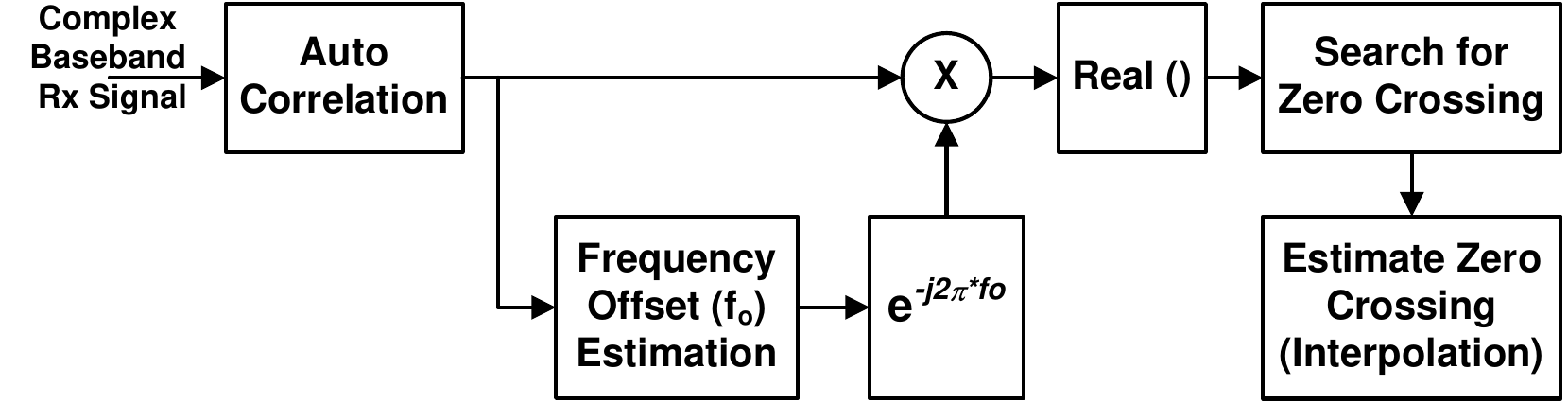}
\caption{Block diagram for the introduced blind symbol rate estimator}
\label{Fig2}
\end{figure}

It can be shown that the time-averaged autocorrelation may be used to estimate the above autocorrelation function averaged over one period, if enough samples are used. Note that we do not need to know the period when computing the time-averaged autocorrelation. Hence, for the example pulse shape in this paper, the time-averaged autocorrelation of the noiseless received signal is the RC pulse. It is also worth mentioning that if the signal is wide-sense stationary, the autocorrelation function, rather than the autocorrelation averaged over a cycle, is again obtained directly from the time-averaged autocorrelation. Then a simple method is used for detecting the first zero-crossing by following the autocorrelation function from its peak until it goes negative. We need to compute the autocorrelation for a number of lags corresponding to the smallest symbol rate, i.e. the largest number of samples, till the first zero crossing. The zero crossing point is then estimated using linear and non-linear interpolation and the first zero crossing occurs at the symbol period T, which is the reciprocal of the symbol rate to be estimated. Fig. \ref{Fig2} shows a simple block diagram for the estimator. The advantages of the algorithm are that it is very simple, requires no knowledge of any system parameters, and produces an estimate of the symbol rate based on a continuum of values rather than selecting from a pre-defined candidate set.
The accuracy may be inferior to some more complicated SREs but the key point is that as long as it delivers acceptable residual error to the time tracking block, the SRE accuracy is irrelevant to overall system performance. Rather, it is the time tracking block which determines overall system performance.

\subsection{Interpolation Technique}
Since the exact zero crossing point is usually not at a sample point, a very important factor in the algorithm is the interpolation technique used to detect the exact zero crossing.
Both linear and cubic spline interpolation techniques are tested. Linear interpolation can give good results when the sampling rate is an integer multiple of the symbol rate or close to that. Since this is usually not the case, using the cubic spline is needed to achieve usable performance. The technique used for the cubic spline interpolation is the not-a-knot spline interpolator discussed in \cite{Deboor}. Not-a-knot is a method for prescribing the end conditions of the cubic spline polynomial by ensuring third derivative continuity. In the following results we show a comparison between cubic spline and linear interpolation at different conditions.

\subsection{Points of Interpolation}

This factor is related to the above factor, but due to its importance we mentioned in a separate point. Two factors appear here,
\begin{itemize}
\item Number of points:

The number of points is limited by the highest symbol rate (lowest oversampling ratio) to ensure the interpolation points on the same slope. The minimum number of samples per symbol is 8 and by experiment, we found that 5 points are enough for interpolation. This also simplifies the algorithm.

\item Position of points:

The position is also limited by the highest symbol rate for the same previous reason. We tried 3 points before zero crossing and 2 after, and 4 points before and 1 after. The latter gives better performance.

\end{itemize}

\subsection{Frequency Offset Compensation}

From \ref{syseqn}, we notice that resultant RC is rotated with phase shifts corresponding to the carrier frequency offset. In our algorithm we estimate the frequency offset from the angle of R(1),
\begin{equation}
    \angle{R(1)}\ =\ 2\pi f_o
\end{equation}

\section{Improvements for the Current Algorithm}

\subsection{Using Other Zero Crossings}

The first improvement is to consider other zero crossings. The enhancement expected from the farther ZCs resulting from dividing the interpolation error by the zero crossing number. Hence, the higher the zero crossing, the lower the estimation error.

\subsection{Different Zero Crossings Combining}

This enhancement is based on the previous one. It is a way to make use of all zero crossings in the same time using proper weighted combining.

One way to select the weights, is to use the SNR at each zero crossing. The SNR at each zero crossing is strongly related to the slope at each zero crossing as the noise is the same on all of them. The derivative of the Raised Cosine is shown in equation \ref{Derivative}.

\begin{dmath}\label{Derivative}
g^{'}\ =\ \frac{\alpha{}\, \sin\!\left(\frac{\pi\, t}{T}\right)\, \sin\!\left(\frac{\pi\, \alpha{}\, t}{T}\right)}{t\, \left(\frac{4\, {\alpha{}}^2\, t^2}{T^2} - 1\right)} - \frac{\cos\!\left(\frac{\pi\, t}{T}\right)\, \cos\!\left(\frac{\pi\, \alpha{}\, t}{T}\right)}{t\, \left(\frac{4\, {\alpha{}}^2\, t^2}{T^2} - 1\right)} + \frac{T\, \sin\!\left(\frac{\pi\, t}{T}\right)\, \cos\!\left(\frac{\pi\, \alpha{}\, t}{T}\right)}{\pi\, t^2\, \left(\frac{4\, {\alpha{}}^2\, t^2}{T^2} - 1\right)} + \frac{8\, {\alpha{}}^2\, \sin\!\left(\frac{\pi\, t}{T}\right)\, \cos\!\left(\frac{\pi\, \alpha{}\, t}{T}\right)}{\pi\, T\, {\left(\frac{4\, {\alpha{}}^2\, t^2}{T^2} - 1\right)}^2}
\end{dmath}

We can Note that the derivative disappears for all terms except the second term, except for the special case of $\alpha=\frac{1}{2}$ (do using L'hopital). Moreover, the result is easily separated into $T$ in the denominator and a term that is a function of $\alpha$ and the zero crossing number (call it $m$) in the numerator as in equation \ref{Derivative_Reduced}.

\begin{equation}\label{Derivative_Reduced}
g^{'}(mT)\ =\ \frac{\frac{1}{T}\ (-1)^{(m+1)}\ cos(\pi \alpha m)}{[m(4 \alpha^2 m^2-1)]}
\end{equation}

This means that the ratio of $|g^{'}(mT)|^2$ is independent of T since you will divide by the sum of ratios.

We need to know the roll-off to use this method, however, which is ok for our application (An interesting side product of this method is that we can estimate roll-off for uknown modulations). Still, we can use the same weights by computing them online from estimated Raised Cosine pulse assuming linear relation ship around the zero crossing for simplicity.

We have to recall here that we a performance trade-off on the different zero crossing. So far, we consider the effect of the noise on the different zero crossings and as we expect if we consider this factor only, the first zero crossing will dominate the results. However, the other important factor is the enhancements from the error reduction due to dividing by the zero crossing number. For now, to take this into consideration, the previous weights may be scaled by the square of the zero crossing number $m$ in the numerator. In conclusion, the final weights are as in equation \ref{Weights}, for each zero crossing. Finally, the final estimated symbol rate is the sum of all weights multiplied by zero crossings, then normalized by the sum of all weights as shown in equation \ref{Final_Estimate}. The final weights are summerized in table II.

\begin{equation}\label{Weights}
W(m)\ =\ \bigg[\frac{cos(\pi \alpha m)}{4 \alpha^2 m^2-1}\bigg]^2
\end{equation}

\begin{equation}\label{Final_Estimate}
\bar{z}\ =\ \frac{\sum\limits_{m=1}^p \tilde{z}(m) W(m)}{\sum\limits_{m=1}^p W(m)}
\end{equation}

\begin{center}

\begin{table}
\caption{Different Weights for the Zero Crossing Combining, Weights A: based on slope only (perfect), Weights B: based on slope only (linear interpolation on the single pulse), Weights C: based on slope and zero crossing}

\begin{center}

\begin{tabular}{|l|l|l|l|}
\hline
ZC No. & Weights (A) & Weights (B) & Weights (C)\tabularnewline
\hline
\hline
1 & 0.7440 & 0.7696 & 0.2911\tabularnewline
\hline
2 & 0.1637 & 0.1491 & 0.2561 \tabularnewline
\hline
3 & 0.0585 & 0.0517 & 0.2058 \tabularnewline
\hline
4 & 0.0239 & 0.0209 & 0.1498\tabularnewline
\hline
5 & 0.0099 & 0.0087 & 0.0972\tabularnewline
\hline
\end{tabular}

\end{center}

\end{table}

\label{t2}

\end{center}

\section{SIMULATION RESULTS}
For the following simulations, the sampling frequency is fixed to 56 MHz, i.e. we have 8 samples as minimum until the first zero crossing (corresponds to highest symbol rate 7 MHz) and a maximum of 56 samples (corresponds to lowest symbol rate 1 MHz). The symbol rate ranges from 1 to 7 MSymbols/sec and the QAM constellation order is 256. Roll-off factor is fixed to the one used in the DVB-C standard (0.15) and the SRRC pulse spans 12 symbol periods on each side.
In order to simulate different symbol rates, the transmitted signal is interpolated to a much higher rate (common multiple between symbol rate and sampling rate) then decimated to the required rate.
We measure the performance of our scheme under different conditions including the case when a single symbol is transmitted noiselessly, which represents the best performance we can obtain.
The estimator performance is measured as a normalized root mean square error (NRMSE) except for SNR test where we use normalized mean square error (NMSE) for performance comparison. The NMSE is defined in \ref{NMSE} and the NRMSE is the square root of the NMSE.
\begin{equation}\label{NMSE}
    NMSE = E\bigg[\bigg(\frac{z-\tilde{z}}{z}\bigg)^2\bigg]
\end{equation}

where, z is the perfect zero crossing and $\tilde{z}$ is the estimated zero crossing, E[.] denotes the expectation.
In the following simulations we show the effect of different factors on the estimator performance where the SNR is fixed to 15 dB. For sufficiently long decision time, much lower SNR may be used. For the convenience of shorter run times, however, we use the higher SNR.


\subsection{Symbol Rate Estimator Error Classification}

To better understand the estimator performance limits, we go over all sources of errors in more details. As described in Section III, first step in our proposed algorithm is to estimate the pulse shaping by calculating the autocorrelation of the received signal over a long length in order to average out the additive whites noise and random data effects. The second is step is performing interpolation to find out the zero crossing time. Therefore, we can classify the estimation errors into two main classes: (1) Raised-Cosine pulse estimation error, and (2) Interpolation error. Moreover, to eliminate the Raised-Cosine estimation error and focus only on interpolation error, we generate a single pulse Raised-Cosine, i.e., without convolving with data and without any additional transmit-receive errors including noise, channel or frequency offset. It is hard to completely eliminate the interpolation error. However, it is possible to significantly reduce the interpolation error by significantly increasing the sampling frequency. In the following we go in details over different types of errors and investigate the effect of different system parameters on the estimation error.

\begin{itemize}

\item{Errors in the Estimated Raised-Cosine Pulse}
\begin{itemize}

\item{AWGN and random data averaging: The assumption that $E[{a_n}^2]$ is constant in (3) is only valid for infinite correlation length. However, it should be a good approximation as correlation becomes larger. Similarly, cross correlation between independent identically distributed zero-mean AGWN approaches zero as correlation length becomes large. Hence, increasing correlation length clearly enhances the Raised-Cosine estimation accuracy.}

 Fig. \ref{NMSE_LengthsSamples_True} shows the NRMSE vs. correlation length for 5 and 7 MSymbols/Sec. As the correlation length increases, the estimated Raised Cosine pulse becomes more accurate and the accuracy of symbol estimation increases.

\begin{figure}
\centering
\includegraphics[width=3.2in]{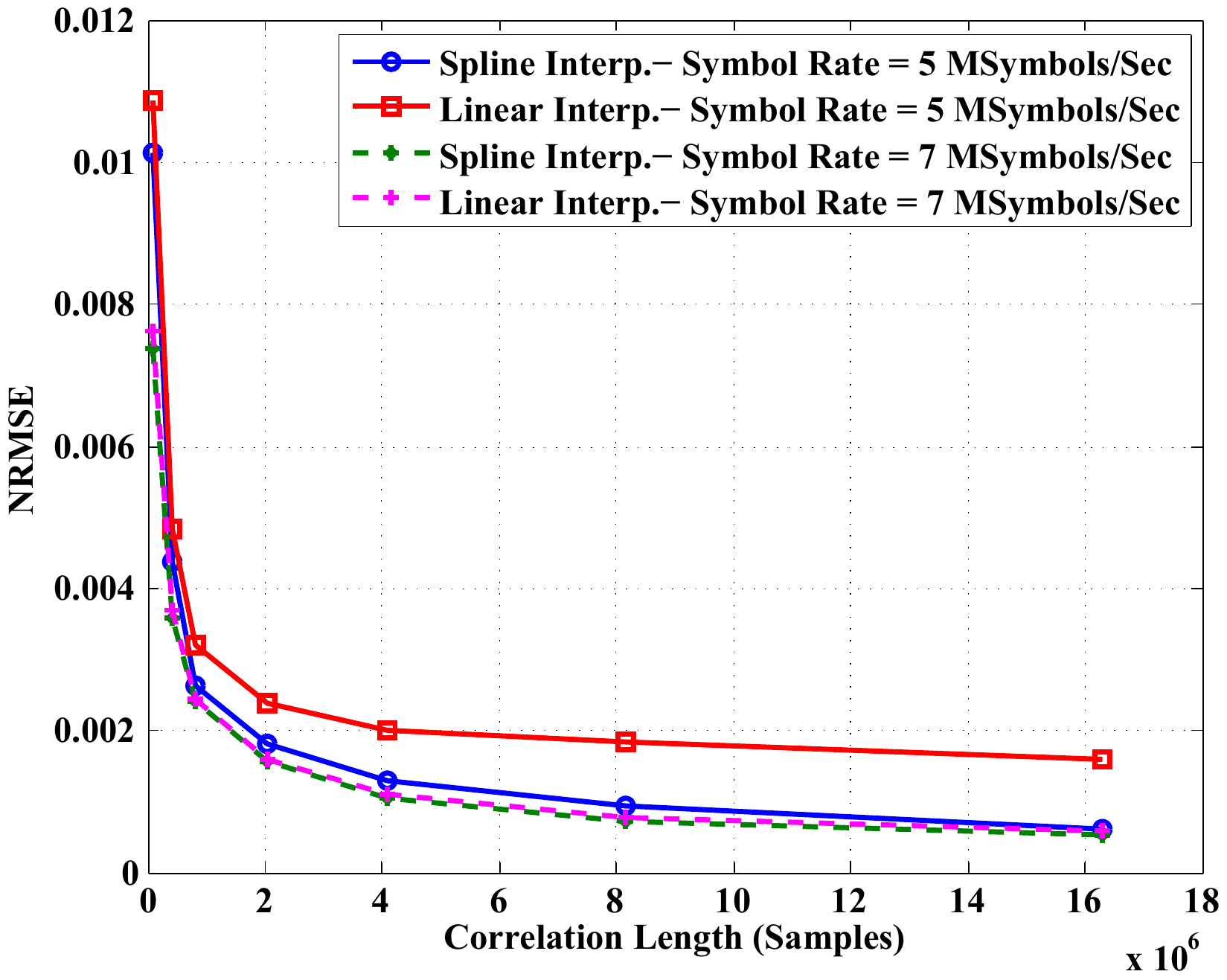}
\caption{NRMSE vs. correlation length, cubic spline and linear interpolation, Symbol rate = 5 and 7 MSymbols/Sec}
\label{NMSE_LengthsSamples_True}
\end{figure}

\item{Truncation error}: In practice, both transmit and receive SRRC filters have limited number of taps while theoretically they have infinite impulse response since they are limited in the frequency domain. Convolving truncated transmit SRRC pulse with itself does not result in a perfect RC pulse even if each SRRC is noise free. As the transmitter filter symbol span increases, zero crossing accuracy increases. It should be noted that the transmit filter symbol span is a design parameter that we have no access to it and is chosen to satisfy the spectral mask requirement of the DVB-C (or other) transmitter. Fig. \ref{NRMSE_FilterSpan} shows the NRMSE for different values of filter span. From the figure we find that a span of 8 symbols is sufficient for our purposes and hence we use it throughout our simulations unless otherwise is mentioned.

\begin{figure}
\centering
\includegraphics[width=3.2in]{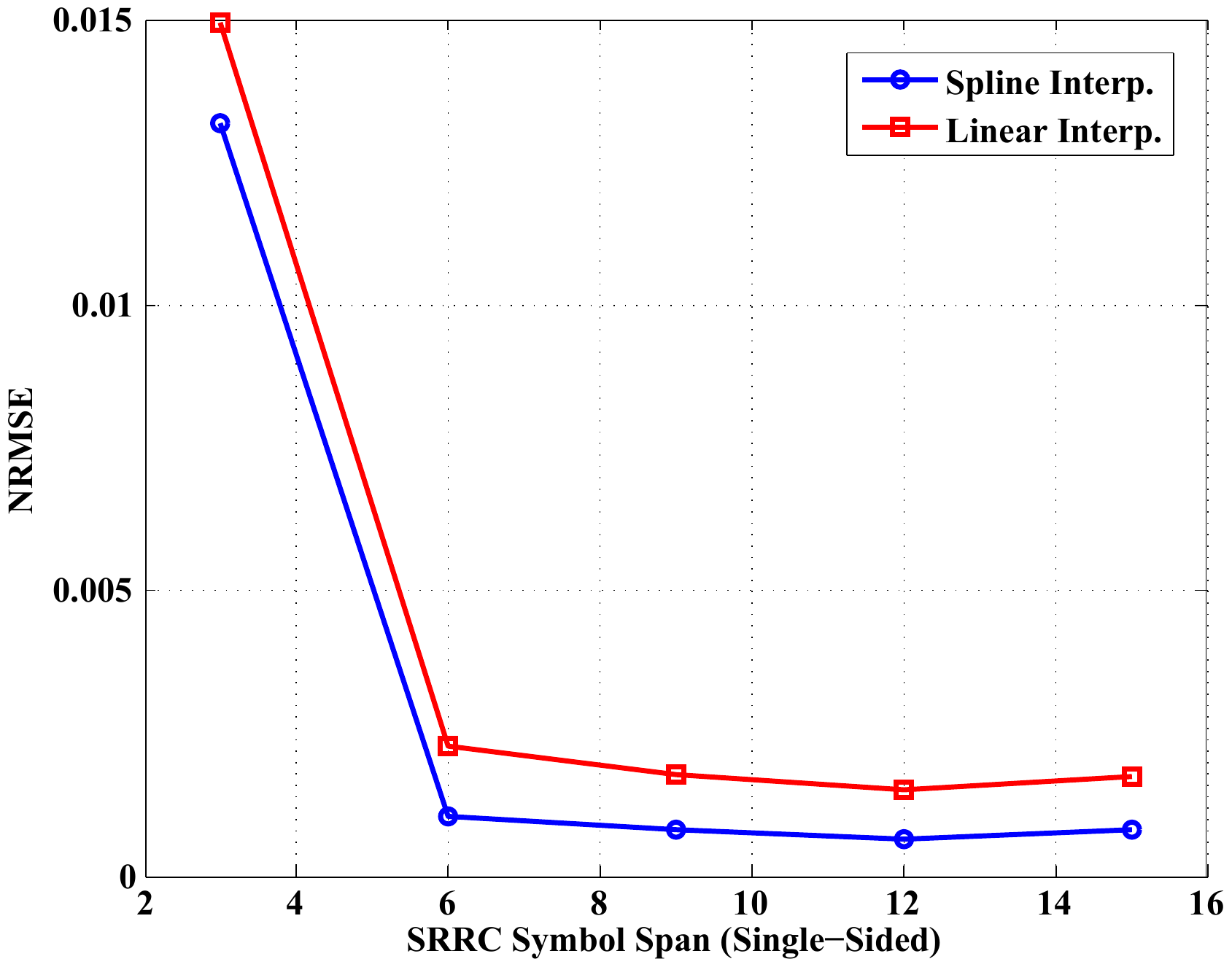}
\caption{NRMSE vs. SRRC symbol span, correlation length = 5 MSamples}
\label{NRMSE_FilterSpan}
\end{figure}



\item{Channel effect (pulse distortion from the multi-path channel, e.g. echo channel): It is clear that multi-path channel will distort the Raised-Cosine pulse causing a shift in the zero-crossing locations. In Fig. \ref{Sinc_Distortion_onside}}, we show the distortion in a single pulse RC caused by different echo channels. In Fig. \ref{AllEchos_ZC1_AllSymbolRates}, we show symbol rate estimation error for the different standardized echo channels at a selected set of the symbol rates spanning from 1 to 7 MSymbols/Sec. It is clear that channel effect causing more degradation at higher symbol rates, i.e., narrower RC pulse. Moreover, echo channel number 3 looks to be the worst case channel conditions for all symbol rates. In Fig. \ref{SinglePulse_withandwithoutEcho}, we show the estimation error for the whole continuous range of symbol rates at the worst case echo channel. As depicted from Fig. \ref{SinglePulse_withandwithoutEcho}, worst case echo channel increased the symbol rate estimation error by ~ 1000 ppm.


\begin{figure}
\centering
\includegraphics[width=3.2in]{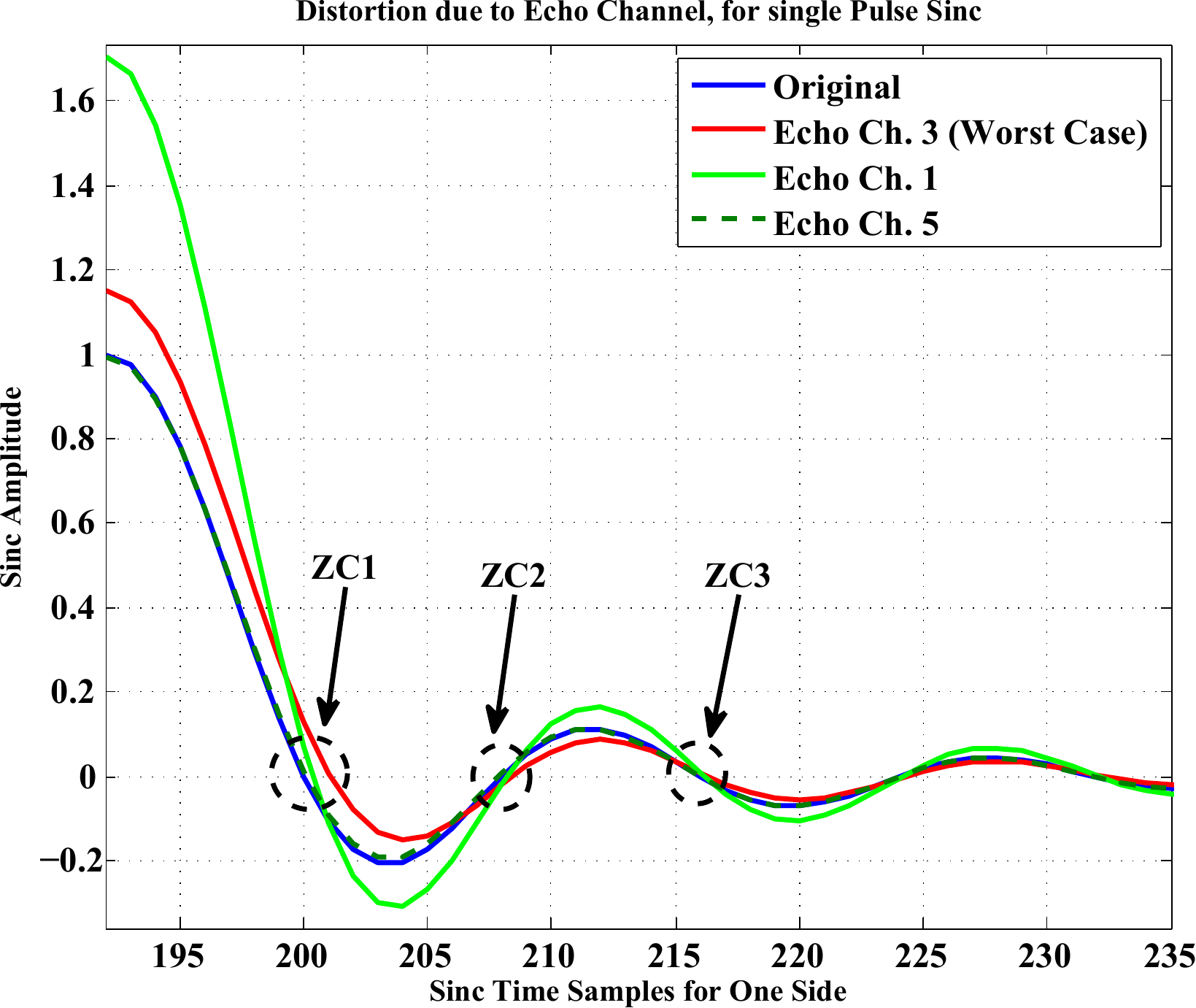}
\caption{Distortion due to Echo Channel, for single Pulse Sinc}
\label{Sinc_Distortion_oneside}
\end{figure}

\begin{figure}
\centering
\includegraphics[width=3.2in]{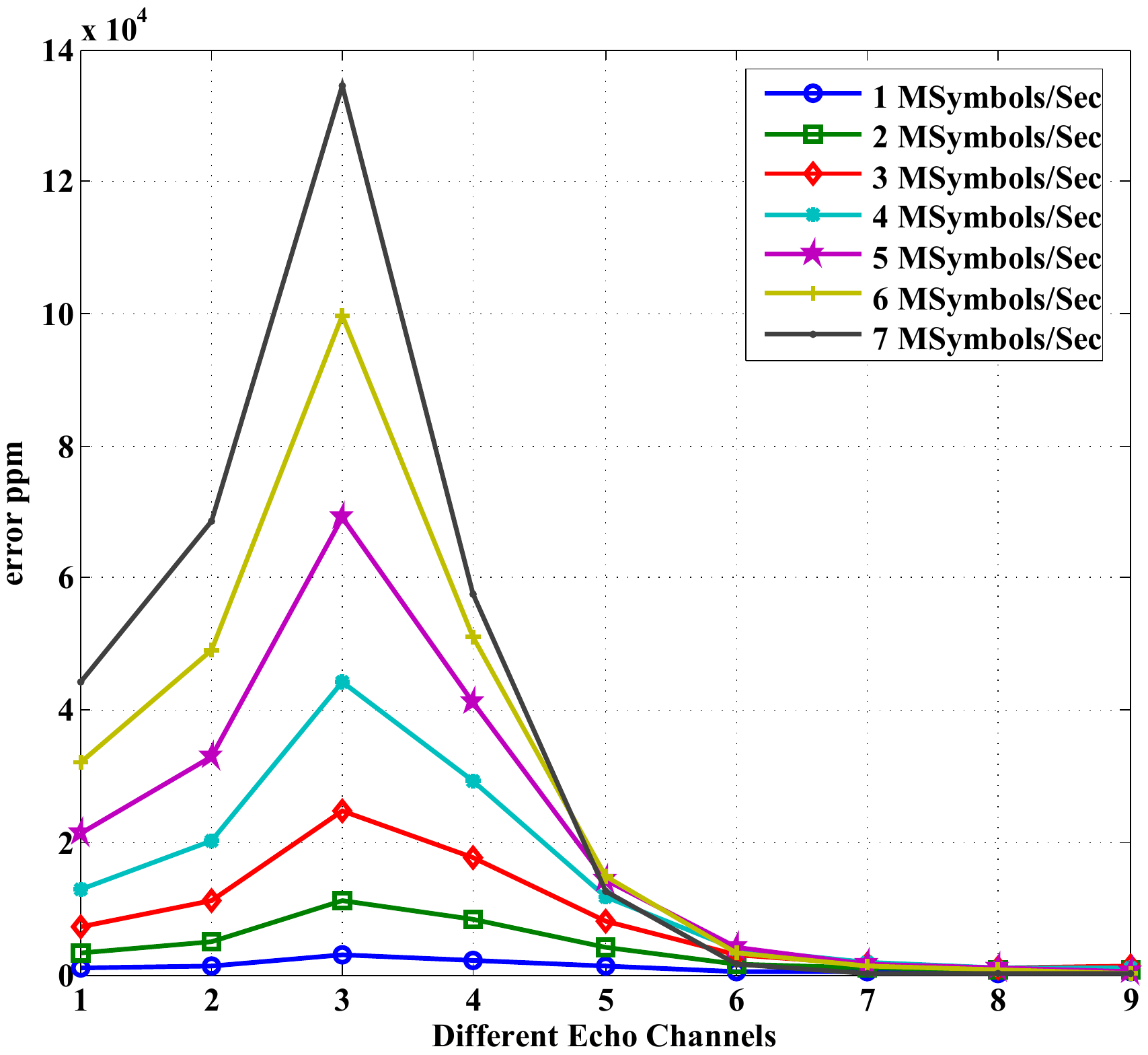}
\caption{Error in ppm for estimated symbol rate in case of single pulse without noise, for all echo channels, different symbol rates (1, 3, 5, 6, 6.58 MSymbols/Sec) for cubic spline interpolation}
\label{AllEchos_ZC1_AllSymbolRates}
\end{figure}

\begin{figure}
\centering
\includegraphics[width=3.2in]{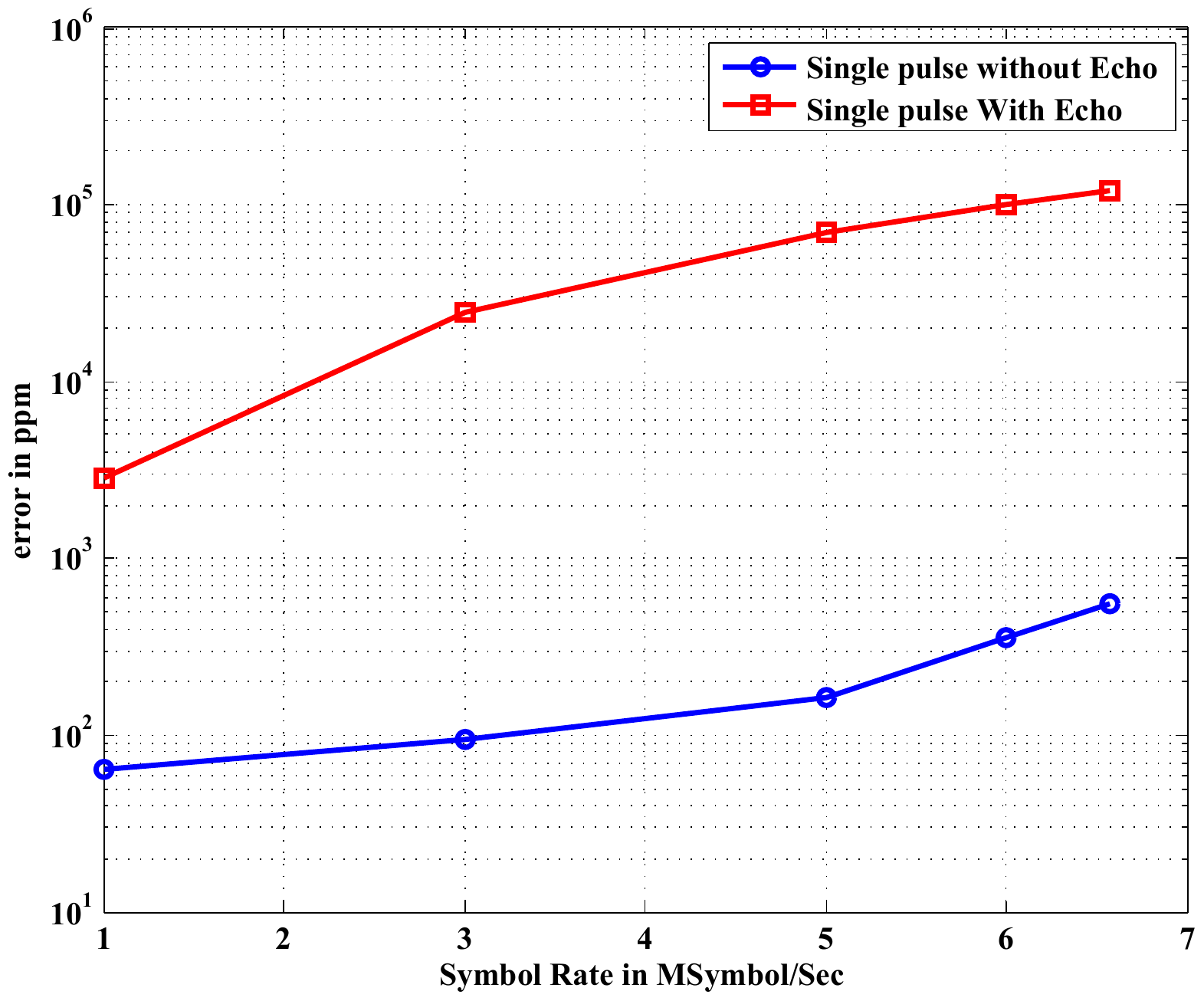}
\caption{Error in ppm for estimated symbol rate in case of single pulse without noise, with and without echo channel (Worst Case), different symbol rates (1, 3, 5, 6, 6.58 MSymbols/Sec) for cubic spline interpolation}
\label{SinglePulse_withandwithoutEcho}
\end{figure}

\item{Frequency Offset Effect}

Fig. \ref{NRMSE_FO} shows the NRMSE at different symbol rates before and after frequency offset compensation. We notice that frequency offset compensation is necessary, especially for low symbol rates (below 2 MSymbols/Sec) as large degradation occurs. This may be explained by the large fraction of bandwidth for a given frequency offset in the case of low symbol rates. The apparent redundancy of frequency offset compensation for high symbol rates is due to the robustness of cyclic-based SREs against frequency offset.

\begin{figure}
\centering
\includegraphics[width=3.2in]{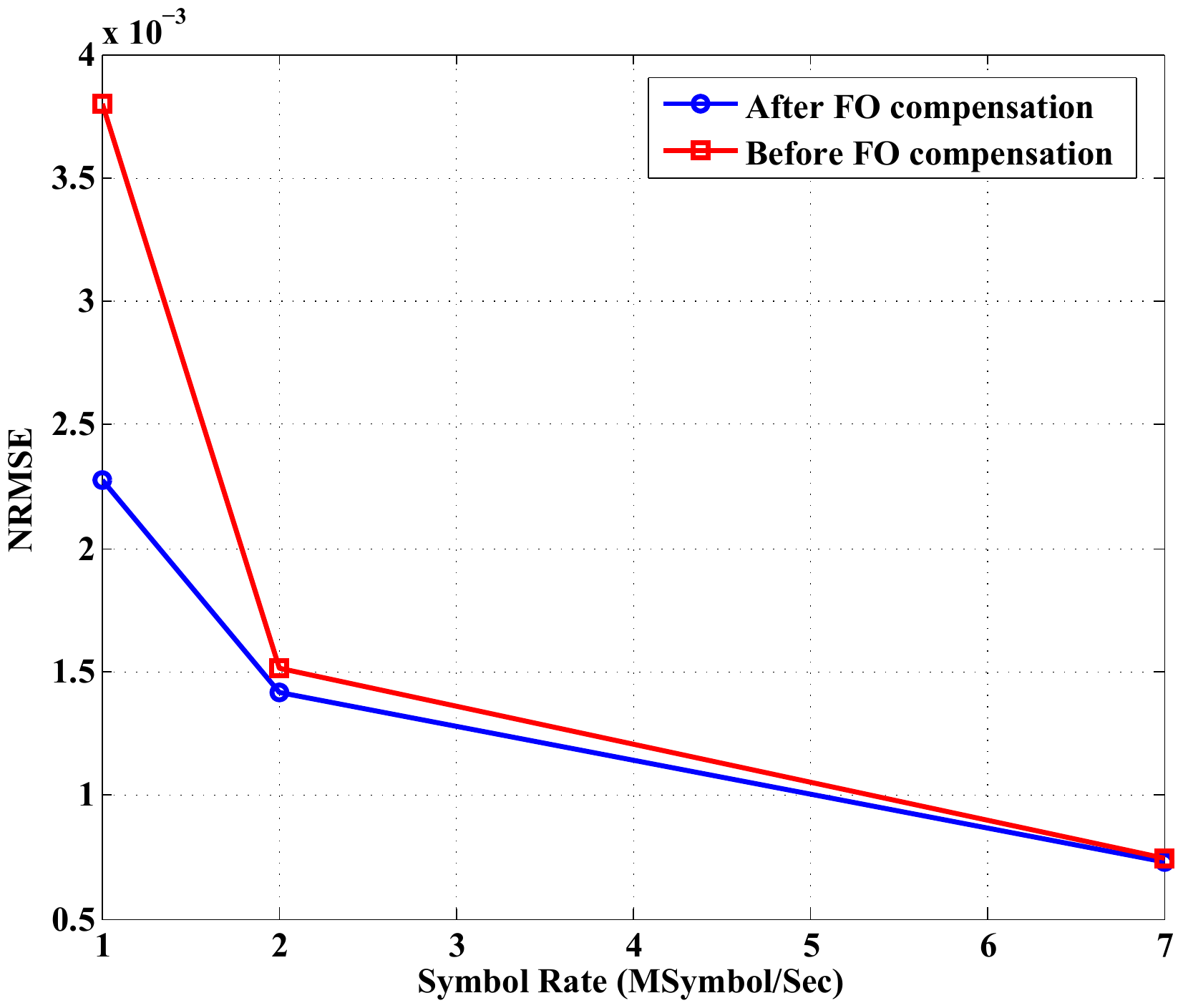}
\caption{NRMSE at different symbol rates (1,2,7 MSymbols/Sec) for cubic spline interpolation, with/without frequency offset compensation, correlation length = 8 M samples, Frequency offset = 150 KHz}
\label{NRMSE_FO}
\end{figure}

\end{itemize}

\item{Interpolation error (calculation of zero crossing)}: Spline interpolation always better than Linear but in some cases the effect of interpolation error is minor (other errors may dominate) and hence using simple Linear interpolation in these cases is better.

For a given sampling rate, as the symbol rate decreases the number of samples per symbol increases and the resolution of the zero crossing increases. On the other hand, the number of symbols being averaged out decreases as the correlation length is fixed. The net result on performance depends on the interpolator type. The linear interpolator benefits more from the increased zero crossing resolution and the performance is close to the cubic spline at low symbol rates while the cubic spline interpolator performs much better at high symbol rates. This is clear from fig. \ref{NRMSE_vsDifferentRates}, where we note the desirable feature of the cubic spline interpolator whose performance is almost constant regardless of the actual symbol rate.
Another factor affecting performance is the ratio of symbol rate to sampling rate. Fig. \ref{NRMSE_Onesymbol_VsALL} shows the NRMSE at different symbol rates covering the entire range ($1 to 7 MSymbols/Sec$) for the limit case (one symbol simulation) for both interpolation types. Apart from the general degradation with increasing symbol rate we notice a variation that depends on the distance between zero crossing and nearest interpolation point. The maximum degradation occurs at the maximum distance (when the zero crossing occurs midway between the nearest two points, i.e., oversampling $\tilde I.5$, where I is an integer number). The worst case is then detected at symbol rate $\tilde 6.5882 MSymbols/Sec$ which equivalent to 8.5 samples per symbol. Again, we note the relative insensitivity of the cubic spline interpolator to symbol rate.
Note that the range $1:7 MSymbols/Sec$ is only for simulating the DVB-C case. However, our estimator does not require any information about symbol rate range and actually can operate on any range of symbol rates such that minimum oversampling is more than 8, which is controlled by sampling frequency.

\begin{figure}
\centering
\includegraphics[width=3.2in]{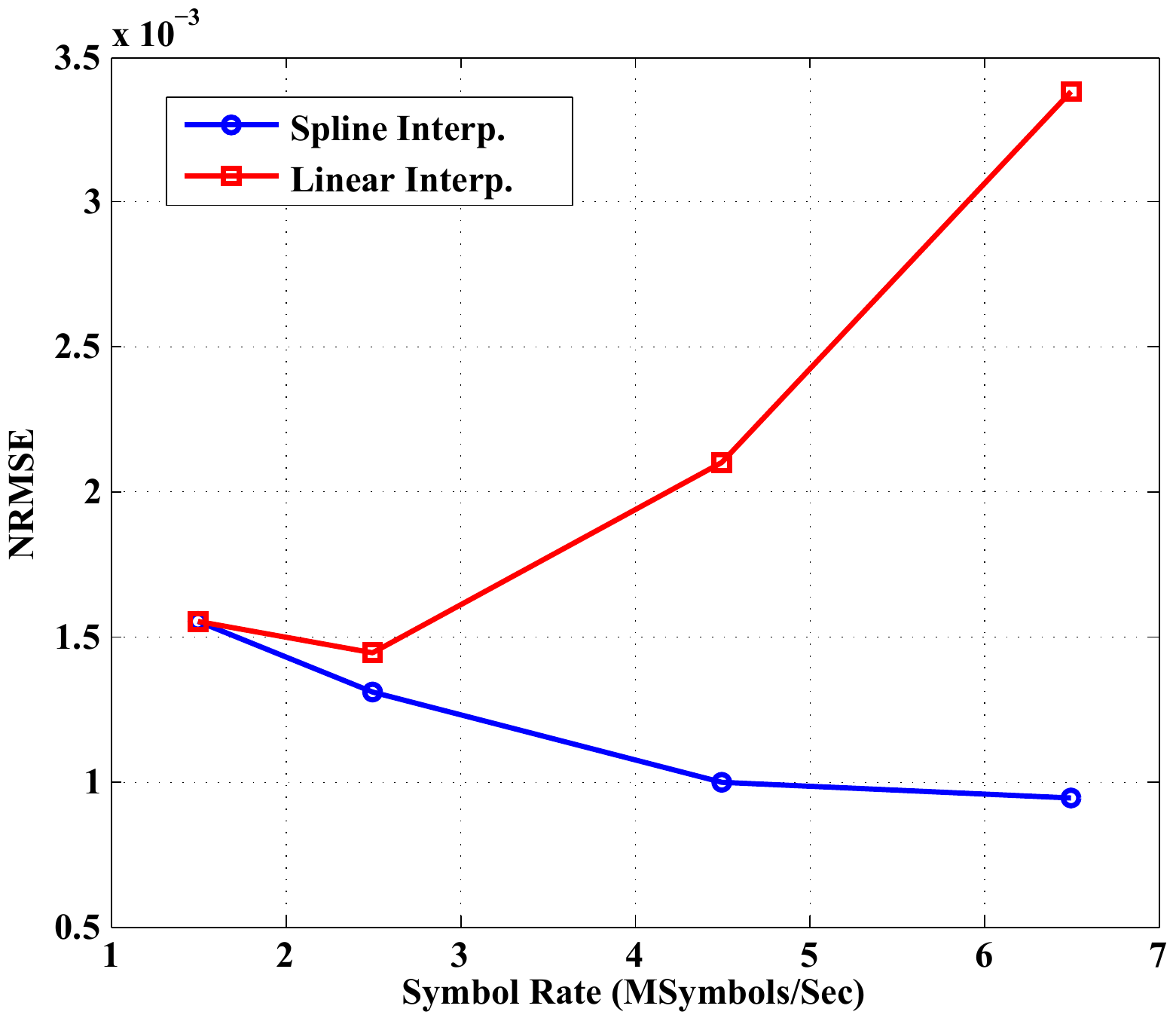}
\caption{NRMSE vs. diffrent symbol rates (1.5, 2.5, 4.5 and 6.5 MSymbols/Sec), correlation length = 8 M samples}
\label{NRMSE_vsDifferentRates}
\end{figure}

\begin{figure}
\centering
\includegraphics[width=3.2in]{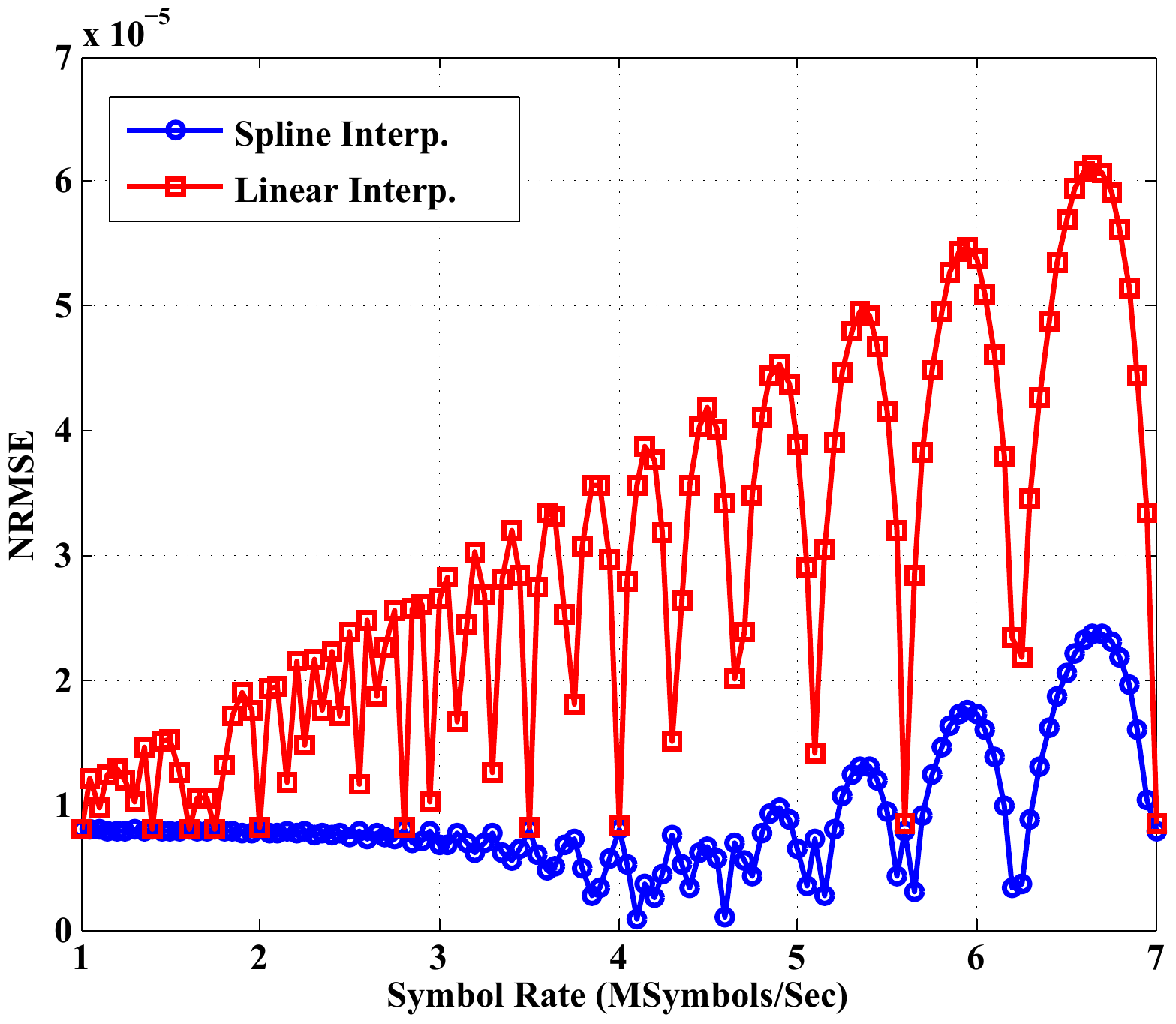}
\caption{NRMSE at different symbol rates for both linear and cubic spline interpolation techniques, One symbol simulation case (limit case)}
\label{NRMSE_Onesymbol_VsALL}
\end{figure}

Another factor that affects the interpolation performance is the characteristics of the pulse, e.g., the roll-off factor of the Raised-Cosine. In general, algorithms that depend on detecting the periodicity of the cyclic autocorrelation are sensitive to the roll-off factor \cite{Dandawate,Mazet}. However, our algorithm performance, appreciably, is not affected by the roll-off factor. Although the DVB-C system uses fixed roll-off factor equals to 0.15, the current algorithm does not exploit this information. Assuming unknown roll-off factor is very useful for other systems that allow variable roll-off factors like DVB-S2 \cite{Mosquera}. Fig. \ref{NRMSE_RollOff_GeneralPoint} shows the NRMSE against different roll-off factors. We notice that cubic spline interpolator performance is almost constant with roll-off variation, and is superior to that of linear interpolation by about an order of magnitude.

\begin{figure}
\centering
\includegraphics[width=3.2in]{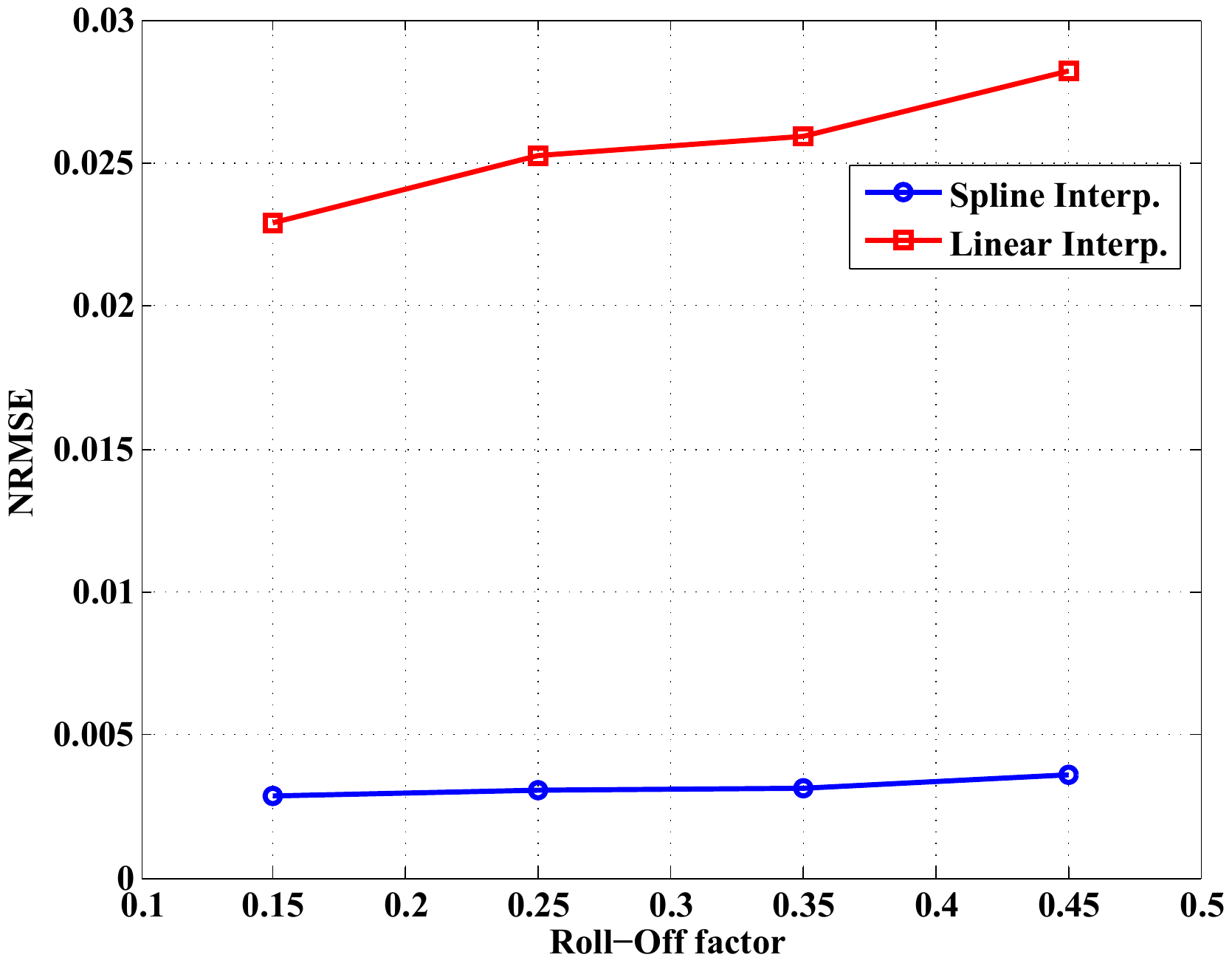}
\caption{NRMSE vs. different Roll-off factors, symbol rate = 5 MSymbols/Sec and correlation length = 5e5 samples}
\label{NRMSE_RollOff_GeneralPoint}
\end{figure}

\end{itemize}

\subsection{First Zero Crossing Performance}

\begin{figure}
\centering
\includegraphics[width=3.2in]{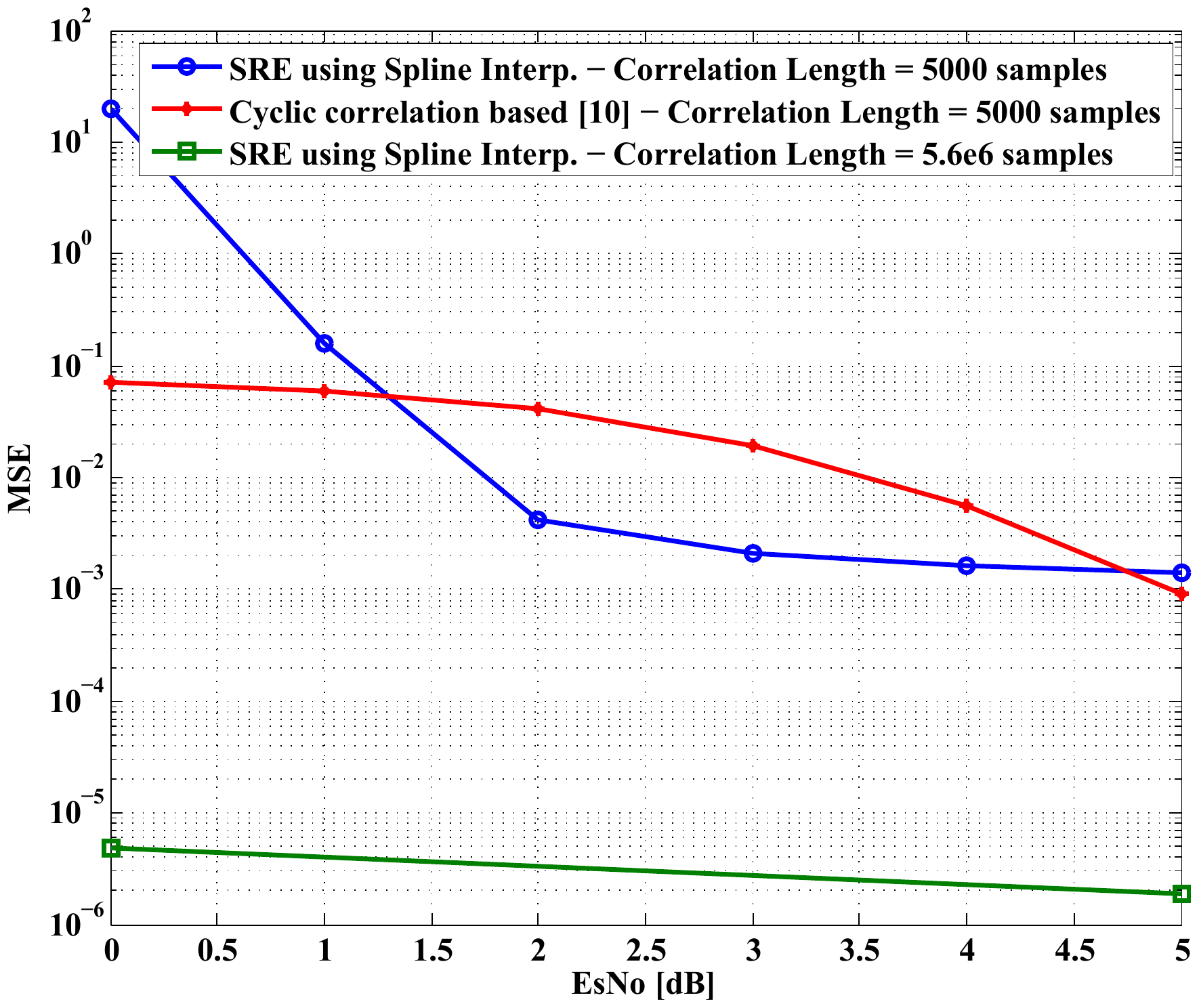}
\caption{NMSE vs. EsNo, cubic spline Interp. based technique at correlation lengths 5000 and 5.6e6 samples vs. cycic correlation based at correlation 5000 sampes, QPSK}
\label{MSE_SNR_All}
\end{figure}

Fig. \ref{MSE_SNR_All} shows the NMSE vs. Es/No for the cubic spline interpolation based SRE compared to cyclic-correlation based SRE shown in \cite{Mosquera}. Note that cyclic-correlation technique search span is $10\%$ around the correct symbol rate and hence the maximum error (at low SNR) does not exceed 0.1. This is not our case where the search span is unknown and the estimator is left free and hence the error is large at low SNR and small correlation lengths. The simulation shows that our performance is better than cyclic-correlation based SRE only at moderate SNR. However, an error floor is expected from the cubic spline based estimator at high SNR. The error floor at small correlation lengths is due to insufficient length to average out data effect. Note that correlation lengths of about 5000 samples are not practical and are shown only for performance comparison.  In practical cases, the correlation length should be increased until the residual error is acceptable to the time tracking algorithm at the lowest operating SNR. For larger observation time, i.e. correlation length, the performance is acceptable as shown in fig. \ref{MSE_SNR_All}. The flooring in that case is due to interpolation accuracy limitation (see fig. \ref{NRMSE_Onesymbol_VsALL}). As mentioned earlier, any improvement beyond what is needed by the time tracking block is unnecessary and has no effect on overall system performance, the determining factor being the accuracy of the time tracking circuit.

For all other simulations, we fix the number of samples of correlation, which corresponds to a certain decision/observation time.

\begin{figure}
\centering
\includegraphics[width=3.2in]{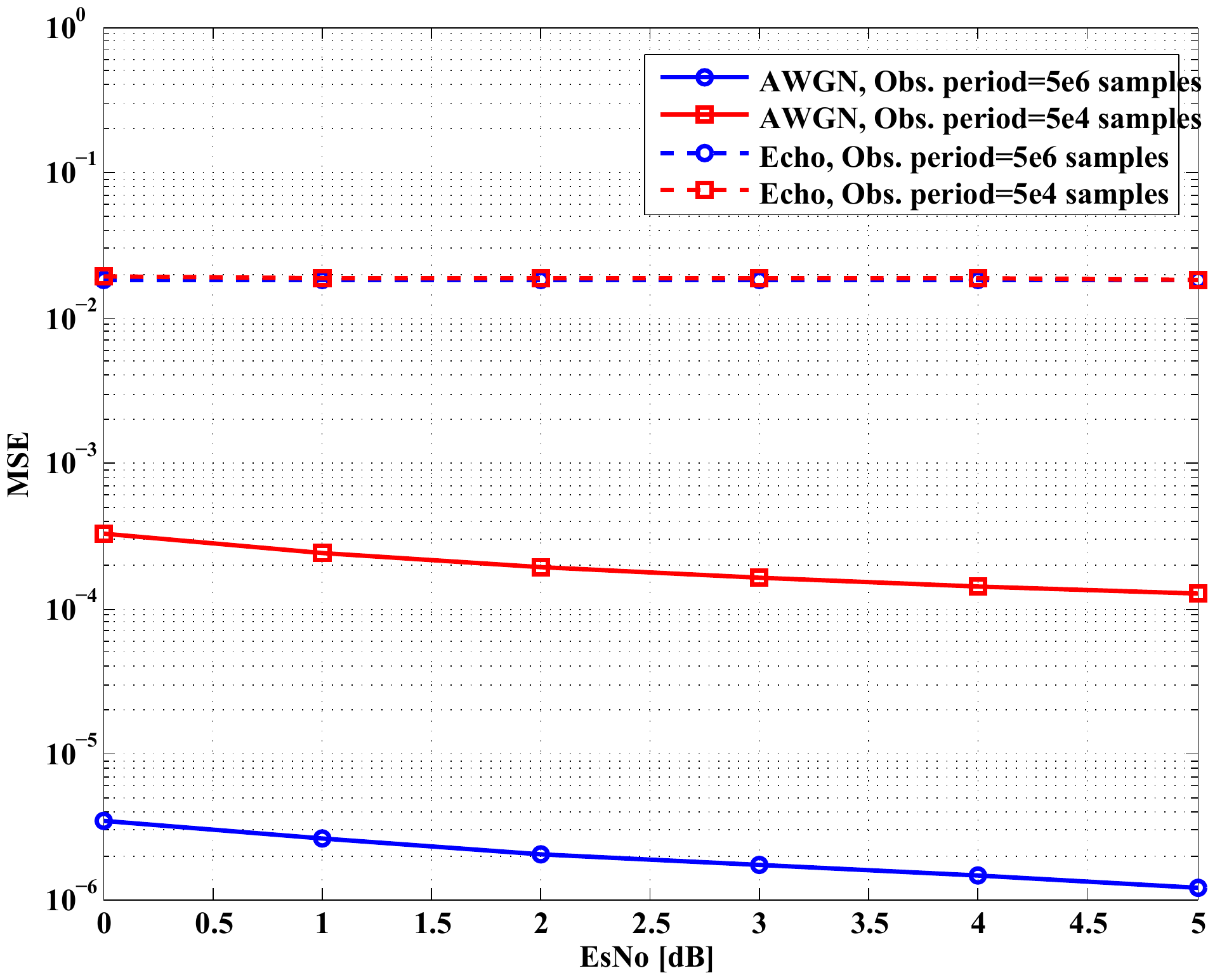}
\caption{MSE for estimated symbol rate, received data with different observation periods, AWGN Vs Fading, Symbol Rate = 7 MSymbols/Sec}
\label{mse_AWGNVsFading}
\end{figure}

In figure \ref{mse_AWGNVsFading}, we show the MSE performance degradation for the symbol rate estimator under echo channel conditions (? which echo) without noise versus AWGN only case. For small observation period, 5e4 samples, MSE for echo channel is worse than AWGN by a factor of 100. The big challenge for echo channel is that symbol rate estimator does not benefit from increasing the observation length. As shown in Fig. \ref{mse_AWGNVsFading}, with increasing observation length from 5e4 to 5e6 samples, MSE for the AWGN case is reduced (enhanced) by a factor of 100, while there is no any improvement for the case of the echo channel. Hence, performance of first zero crossing is not sufficient for proper operation for the time tracking loop and there is no way to enhance the performance with more averaging, i.e., increasing the observation length. This motivates us to investigate other ideas including other zero crossings performance.


\subsection{Considering other Zero Crossings}

\begin{figure}
\centering
\includegraphics[width=3.2in]{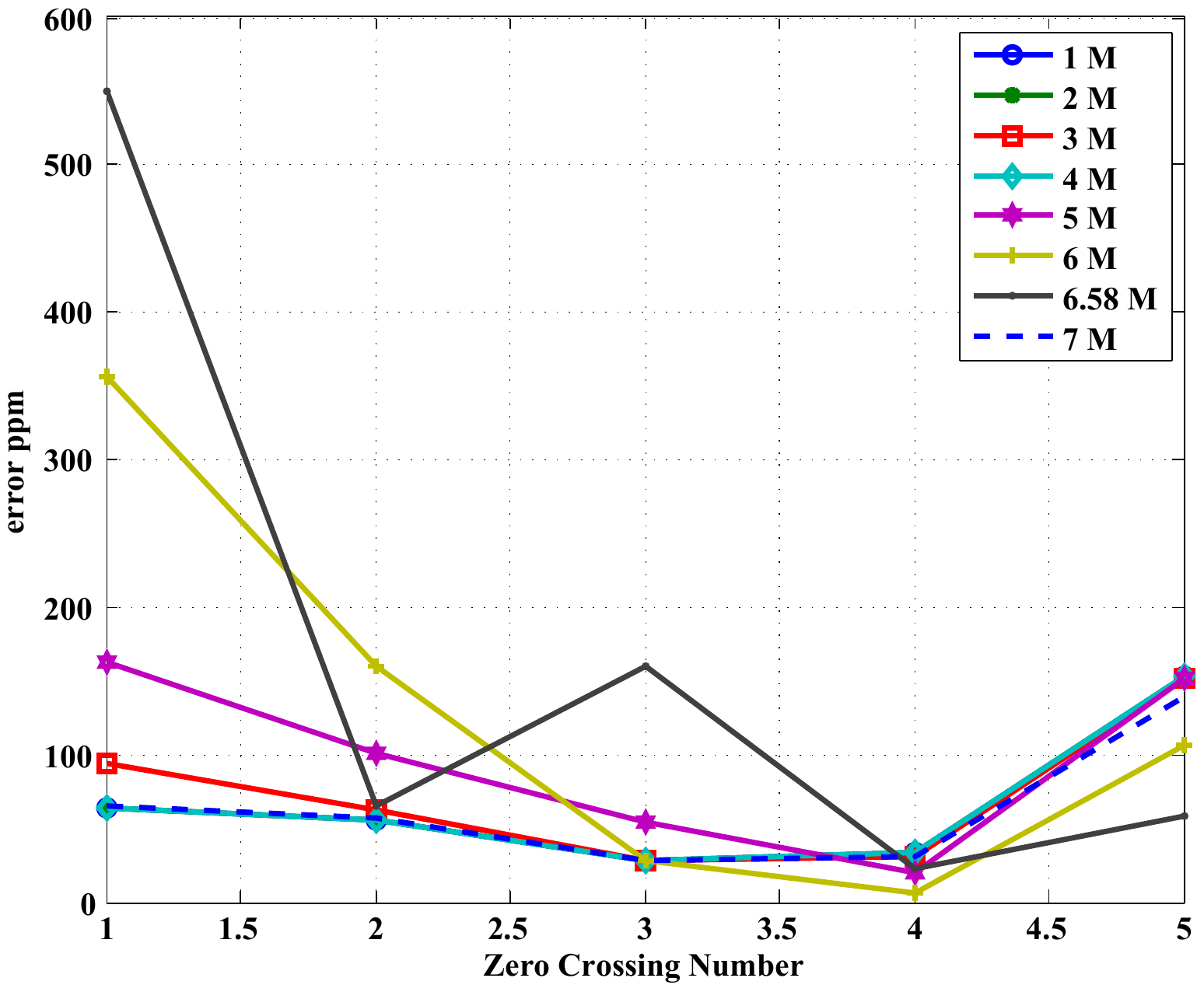}
\caption{Error in ppm for estimated symbol rate in case of single pulse without noise, without channel, different symbol rates (1, 2, 3, 5, 6, 6.58 and 7 MSymbols/Sec), different zero crossings (1:5)}
\label{Without_Echo}
\end{figure}

In the following, we show performance enhancements for the symbol rate estimation algorithm exploiting other zero crossings.
Figure \ref{Without_Echo} shows the error in ppm in case of single pulse (without noise, without channel) for all symbol rates using different zero crossings. This figure shows that performance can be significantly enhanced using other zero crossings. However, performance is not monotonically enhanced with increasing zero crossing. Moreover, the behaviour is different based on the symbol rate. 

\begin{figure*}
    \centering
    \begin{subfigure}[b]{0.45\textwidth}
        \includegraphics[height=2.65in]{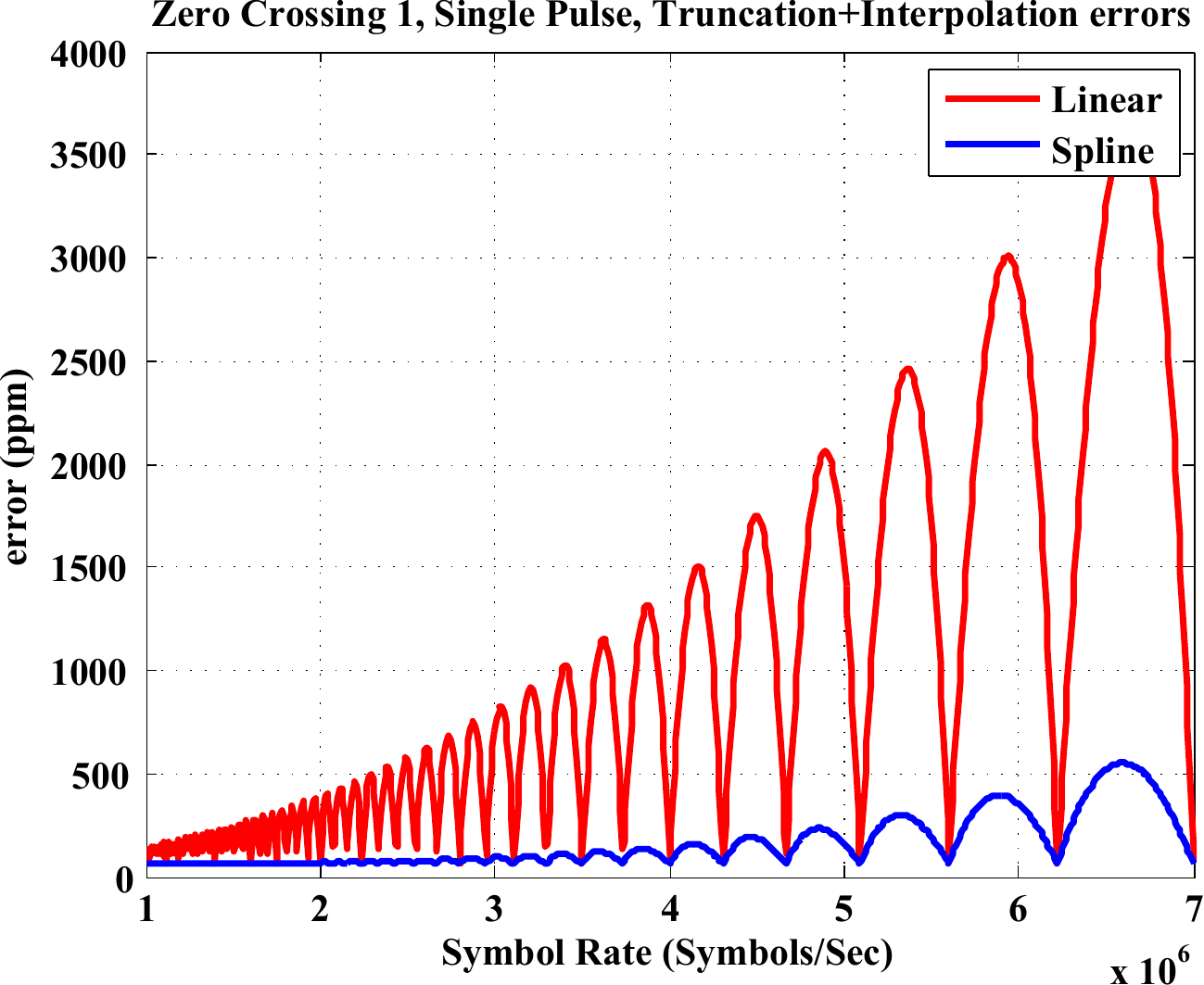}
        \caption{Zeros Crossing One}
    \end{subfigure}%
    ~
    \begin{subfigure}[b]{0.45\textwidth}
        \includegraphics[height=2.65in]{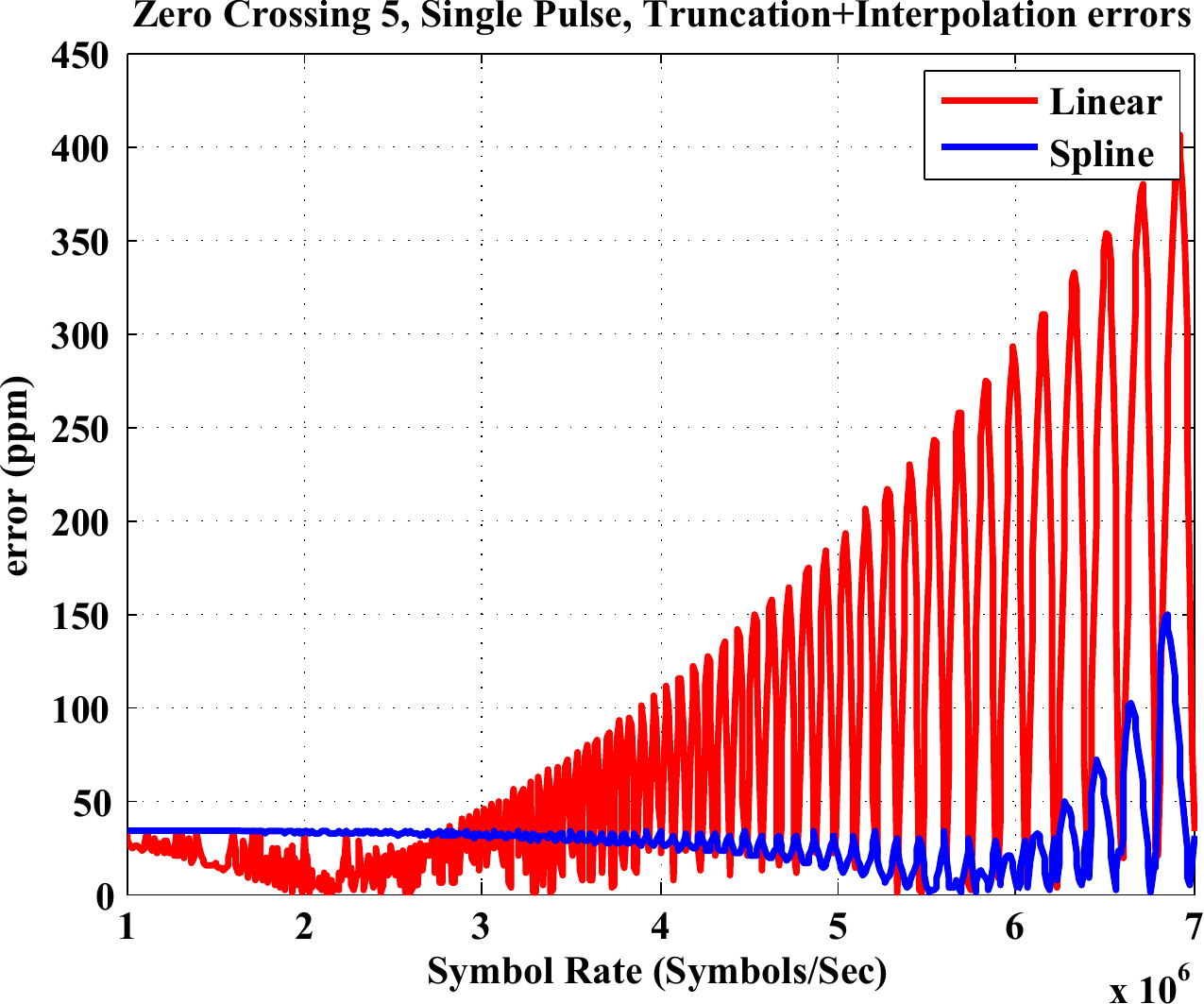}
        \caption{Zeros Crossing Four}
    \end{subfigure}
	~
    
    \begin{subfigure}[b]{0.45\textwidth}
        \includegraphics[height=2.65in]{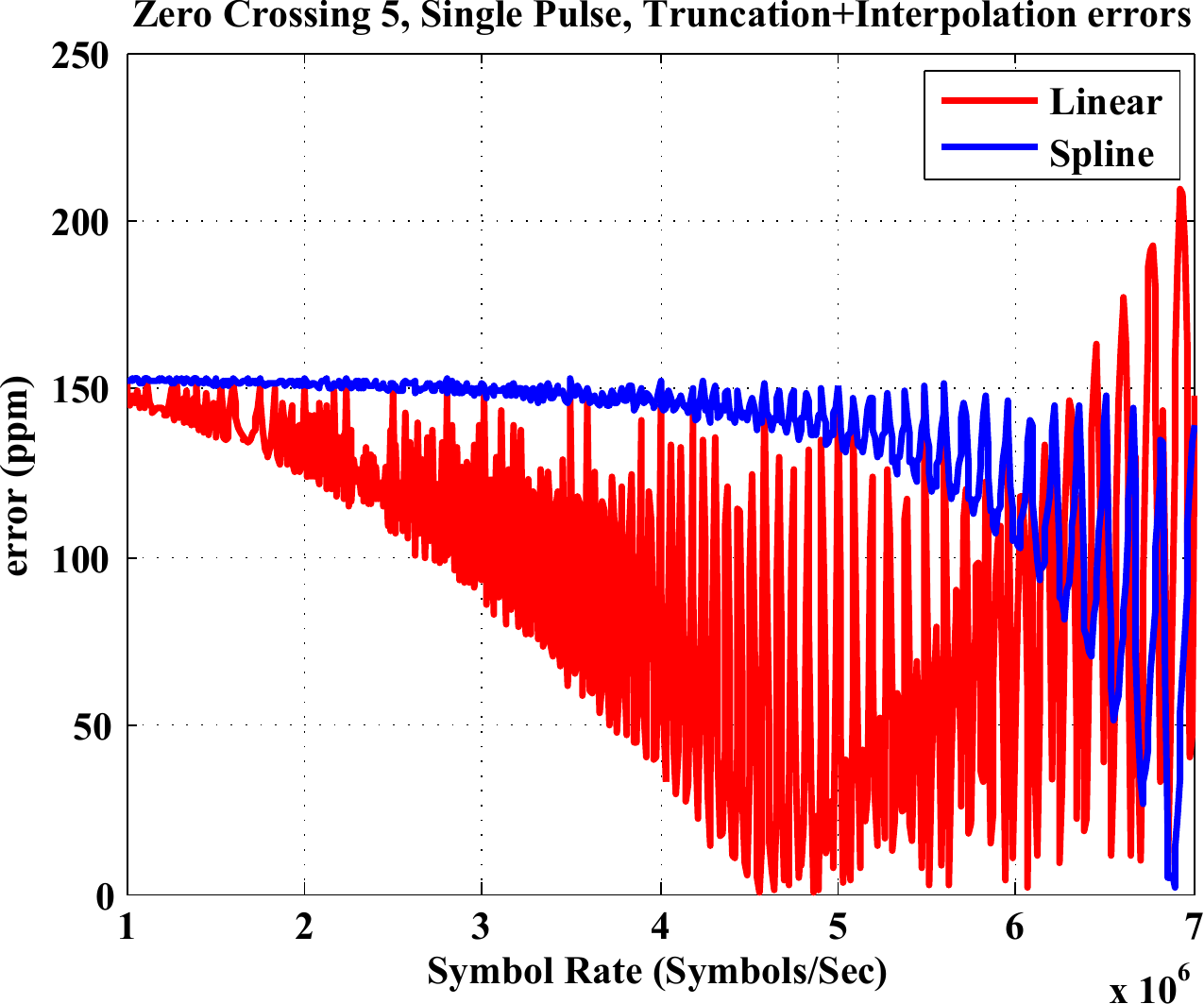}
        \caption{Zeros Crossing Five}
    \end{subfigure}%
    ~
    \begin{subfigure}[b]{0.45\textwidth}
        \includegraphics[height=2.65in]{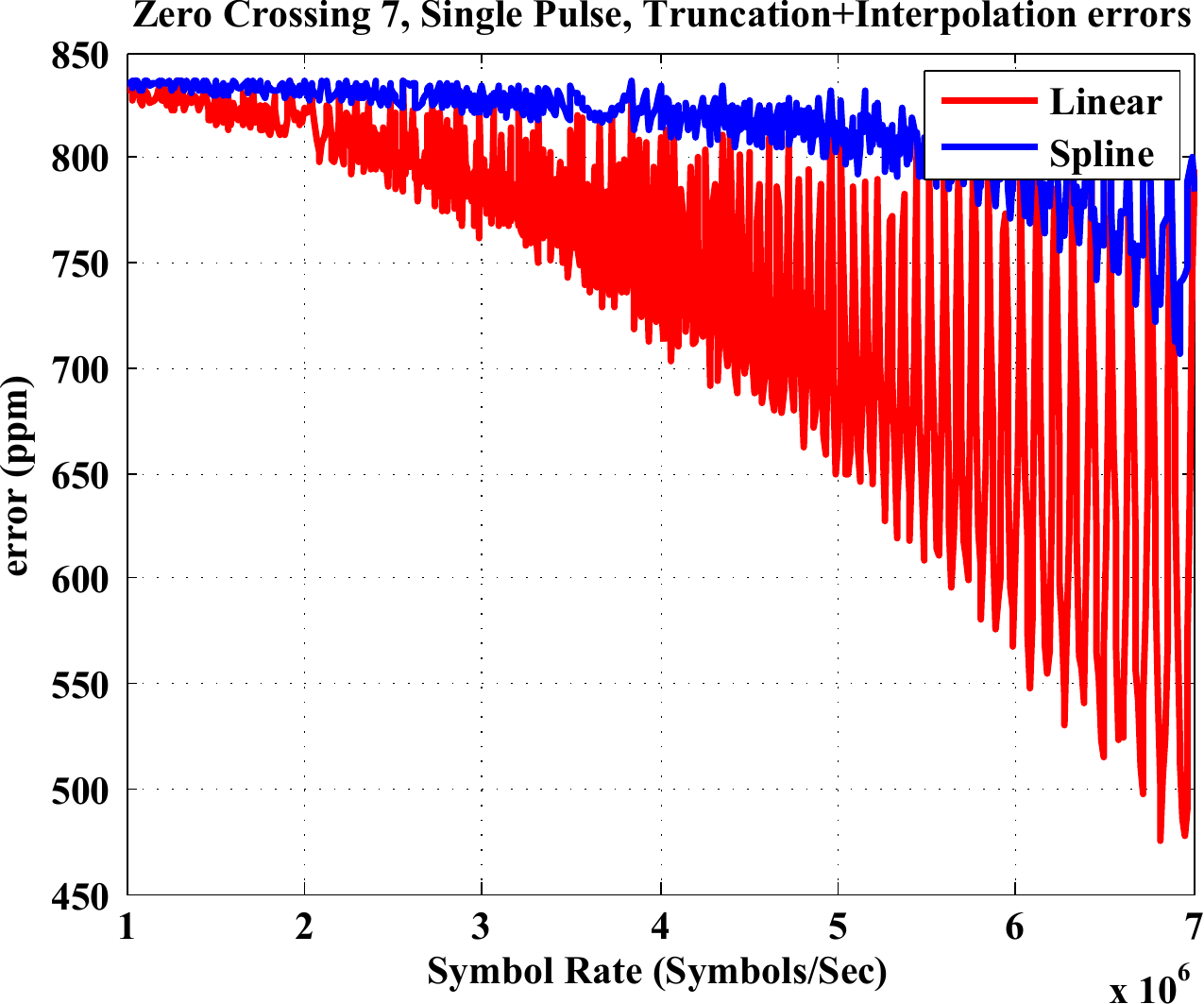}
        \caption{Zeros Crossing Seven}
    \end{subfigure}%
    
        \caption{The total errors in case of single pulse (truncation + interpolation) vs symbol rate, no channel, for different ZCs}
        \label{Total_error_Single_Pulse}
\end{figure*}

To better understand the behaviour of different zero crossings for different symbol rates, we investigate the performance for each zero crossing individually as shown in Fig. \ref{Total_error_Single_Pulse}. Since we investigate the performance of a single pulse, the error (ppm) in Fig. \ref{Total_error_Single_Pulse} represents the combination of the interpolation error and truncation error.

\begin{figure}
\centering
\includegraphics[width=3.2in]{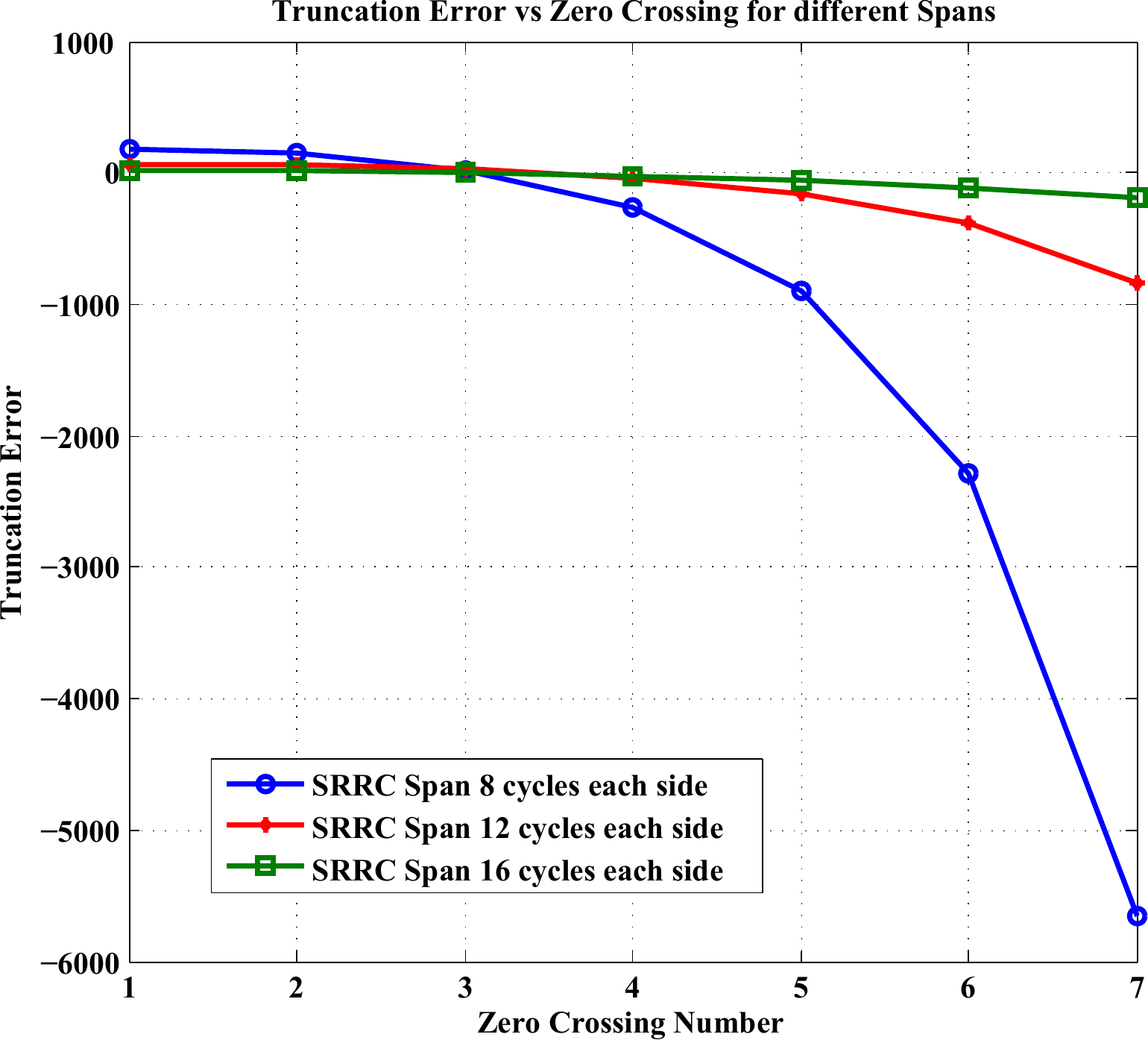}
\caption{Truncation error only in case of single pulse vs ZC number, for different filter spans, any symbol rate}
\label{TruncationError_vs_ZC_DifferentSpans}
\end{figure}

The behaviour is also not clear for two reasons. First, the total error in ppm decreases with increasing the ZC number until ZC5, then increases for ZC6,7. Secondly, the Linear interpolation shows better performance than Spline interpolation for ZC6,7.
To explain this behaviour, we separate the truncation error from the interpolation error. To test the truncation error only (i.e. avoid the interpolation error), we sample the single pulse SRRC at a huge sampling rate (here, we use 560 MHz instead of normal 56 MHz). It is clear here that the truncation error does not change with symbol rate which makes sense as we avoided the interpolation error. In figure \ref{TruncationError_vs_ZC_DifferentSpans}, we show the relation between the truncation error (for any symbol rate, as truncation error doe not change with symbol rate) and the ZC number. The simulation shows that truncation error has a direction (positive and negative) and constant for each ZC number.

\begin{figure}
\centering
\includegraphics[width=3.2in]{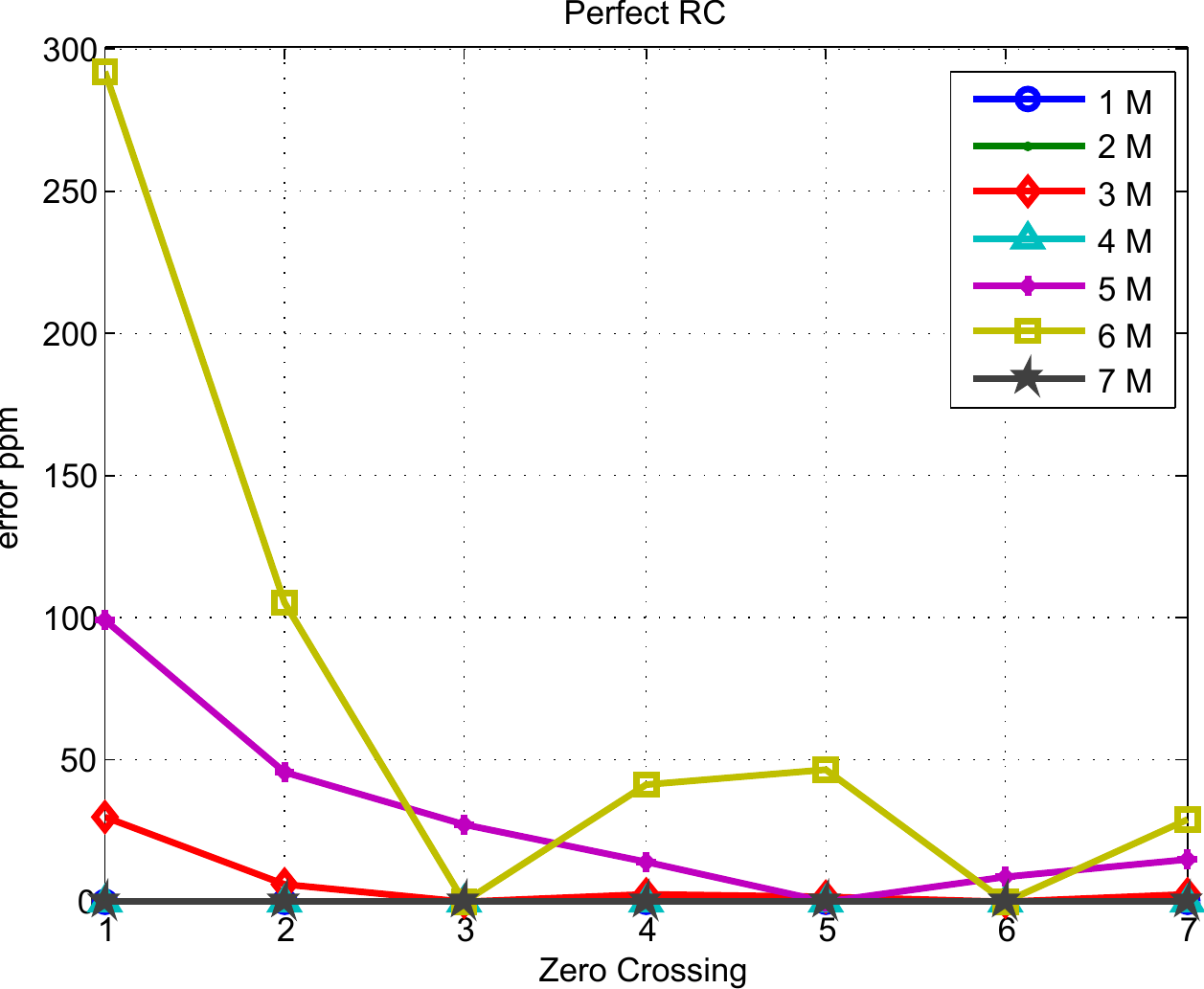}
\caption{Interpolation error only in case of single pulse vs ZC number, for different symbol rates}
\label{All_ZC_PerfectRC}
\end{figure}
\begin{figure*}
    \centering
    \begin{subfigure}[b]{0.5\textwidth}
        \includegraphics[height=2.65in]{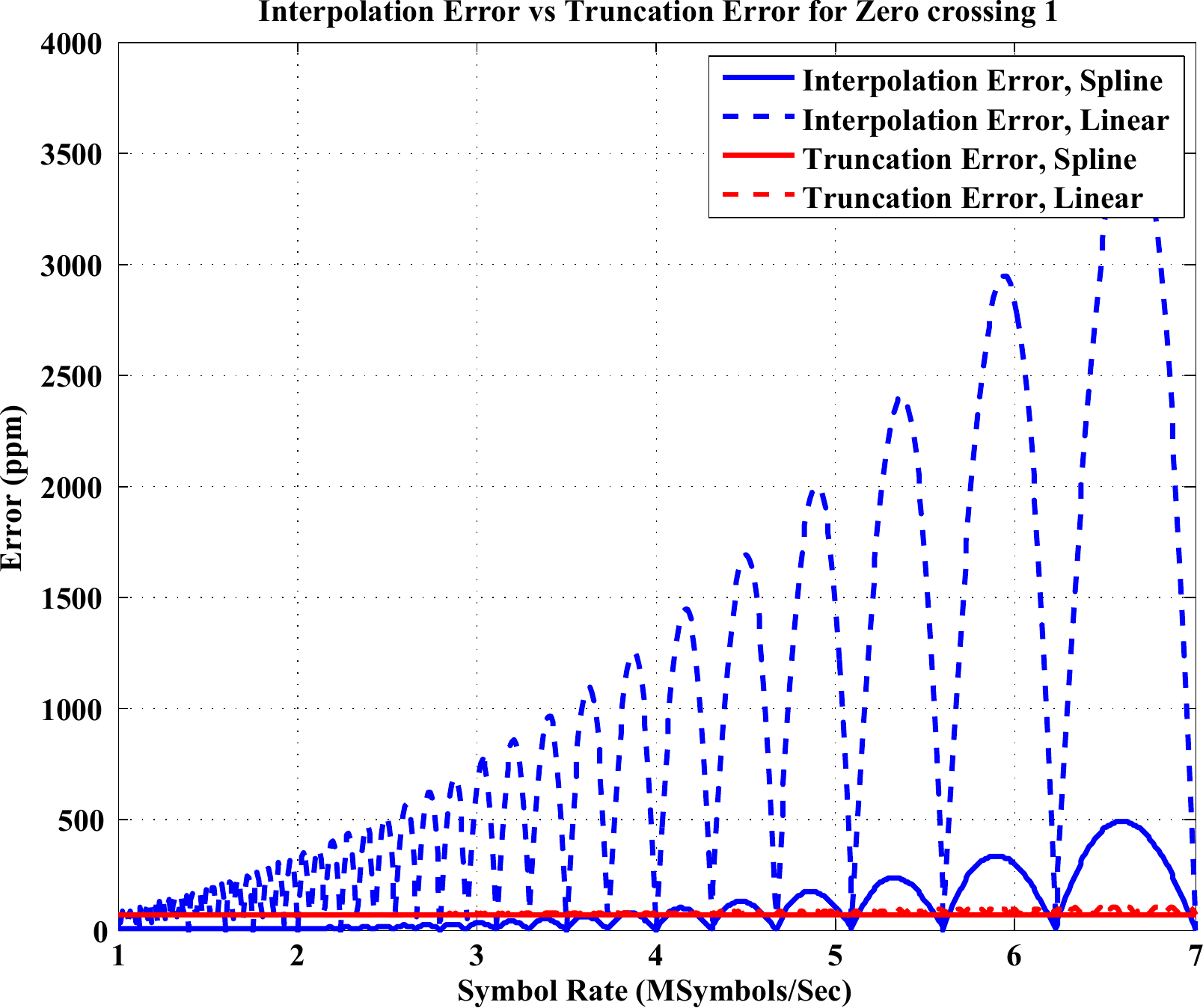}
        \caption{Zero Crossing One}
    \end{subfigure}%
    ~
    \begin{subfigure}[b]{0.5\textwidth}
        \includegraphics[height=2.85in]{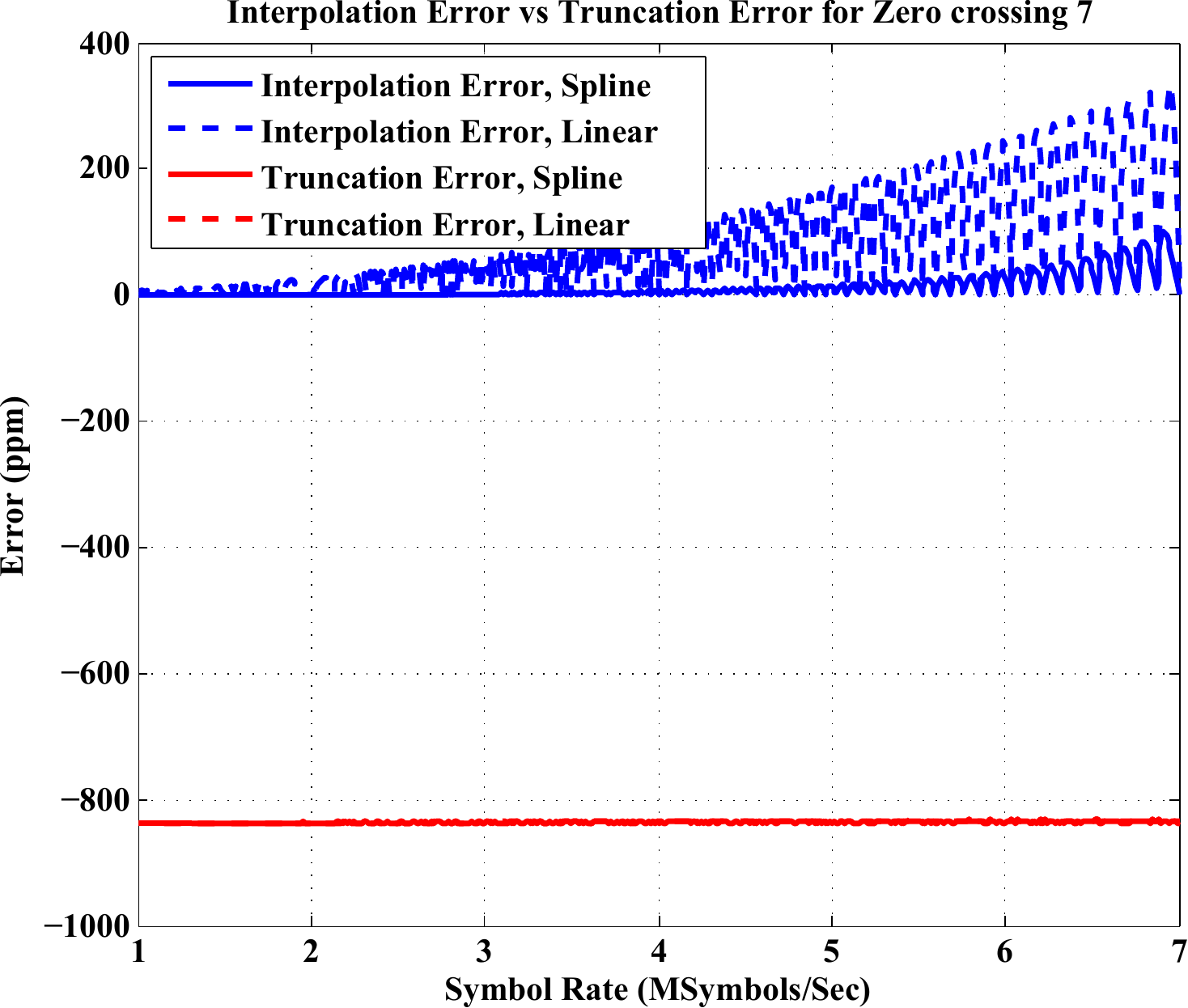}
        \caption{Zero Crossing Seven}
    \end{subfigure}

        \caption{Comparison between truncation error vs interpolation error in case of single pulse, no noise}
        \label{InterpolationError_vsTruncationError_ZeroCrossing}
\end{figure*}

Now, let us check the interpolation error only. To test the interpolation error only, we have to use perfect Sinc pulse (by equation, i.e., not to generate SRRC then autocorrelation). Instead, we increased the SRRC span to a huge number (here, we use 12000 cycle each side). In this case, the error will be absolute zero when the sampling rate is integer multiple of symbol rate as interpolation error will be zero. This is clear in figure \ref{All_ZC_PerfectRC}, which shows the error variation with ZC number for different symbol rates.
When we combine results for truncation error only and interpolation error only, we can understand the behaviour for the estimator in case of AWGN or no noise at all (single pulse). Figure \ref{InterpolationError_vsTruncationError_ZeroCrossing} shows each error alone for both ZC1 and ZC7. Note that interpolation error is always positive. Now, it is clear that the performance degradation for ZC7 is due to the large truncation error. Also, the better performance for Linear interpolation over the Spline interpolation in ZC7 appears as a result of directional truncation error (as large Linear interpolation error in positive compensates with the large truncation error in negative).

\begin{figure}
\centering
\includegraphics[width=3.2in]{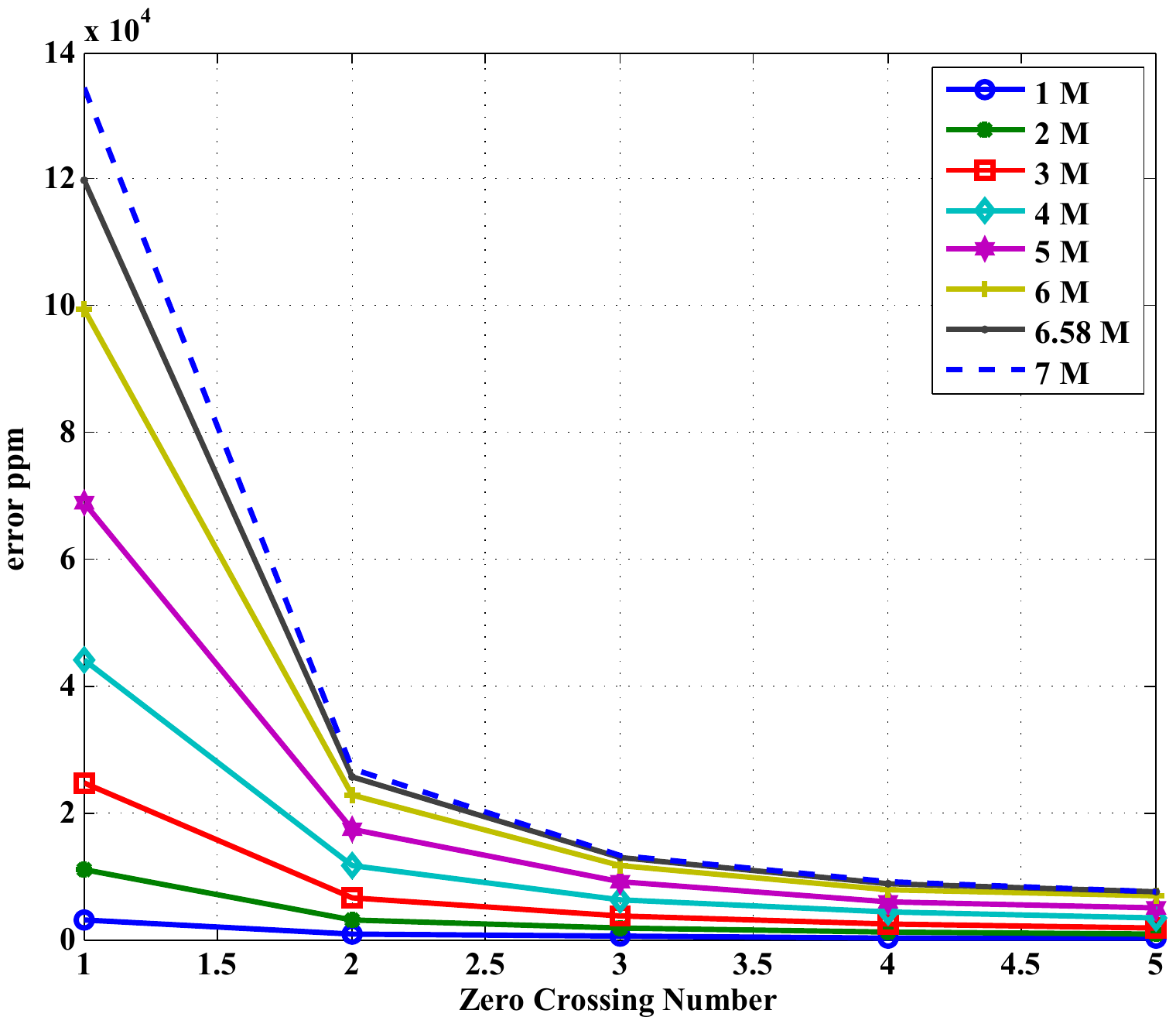}
\caption{Error in ppm for estimated symbol rate in case of single pulse without noise, worst case echo channel, different symbol rates (1, 2, 3, 5, 6, 6.58 and 7 MSymbols/Sec), different zero crossings (1:5)}
\label{WorstCaseEcho_ZC_1to_7}
\end{figure}

\begin{figure}
\centering
\includegraphics[width=3.2in]{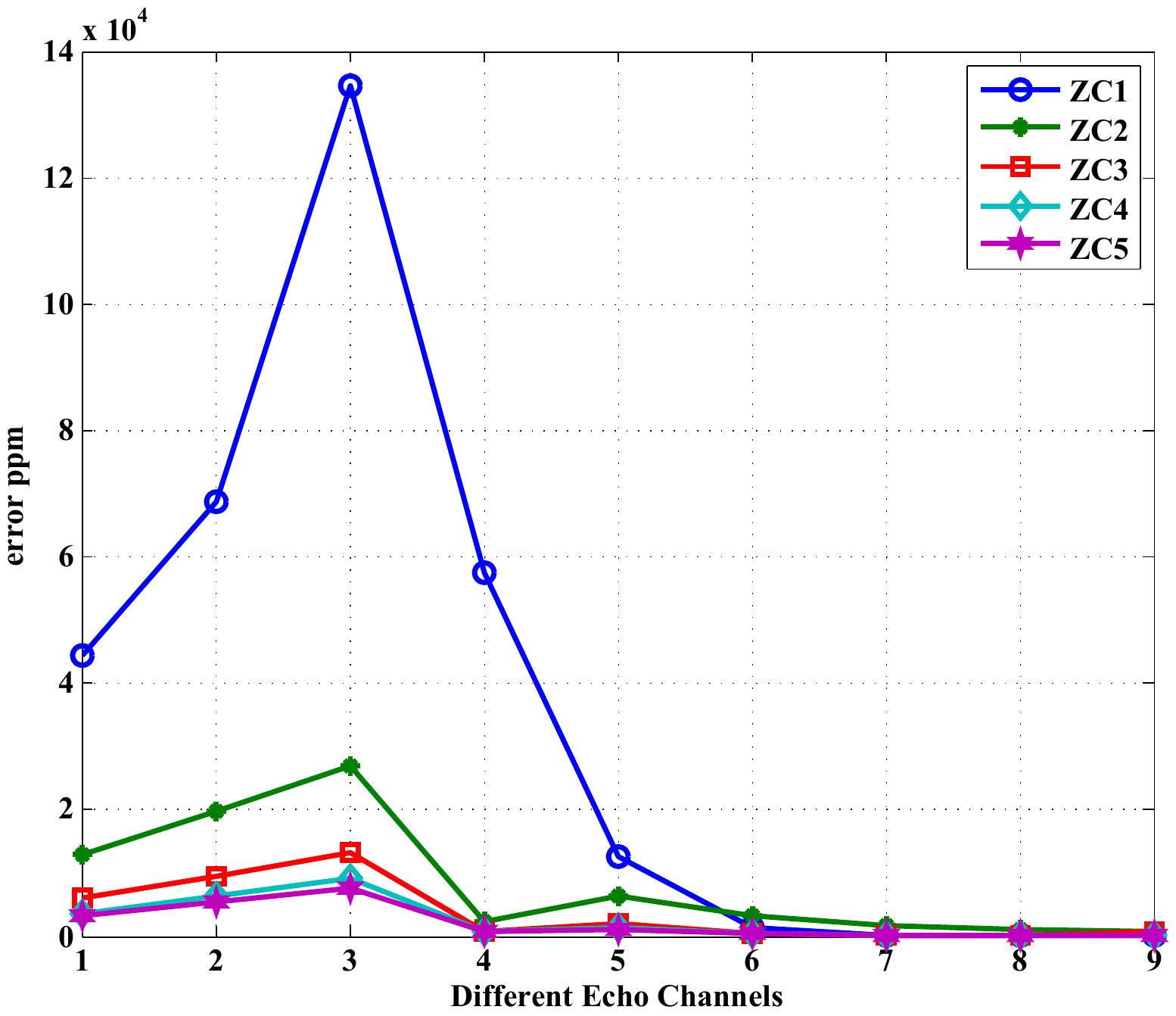}
\caption{Error in ppm for estimated symbol rate in case of single pulse without noise, All standard echo channels, symbol rate = 7 MSymbols/Sec), different zero crossings (1:5)}
\label{AllEchos_ZC}
\end{figure}

For the case of echo channel, the distortion occurred in the RC pulse dominates both truncation and interpolation error (in case of single pulse). In figure \ref{WorstCaseEcho_ZC_1to_7}, we show the performance of the estimator for a single pulse but for the worst case echo channel. Simulation results show significant improvement for the farther zero crossings in the case of echo channel. Then, farther ZC always gives better performance as the distortion error is divided. The performance enhancement for all echo channels are shown in figure \ref{AllEchos_ZC}. 

The conclusion from the previous results is that ZC4 or ZC5 is little better than others in case of no channel (as in figure \ref{Without_Echo}). While, farther ZCs is always much better than earlier in case of channel (as in figure \ref{WorstCaseEcho_ZC_1to_7}).

\begin{figure}
\centering
\includegraphics[width=3.2in]{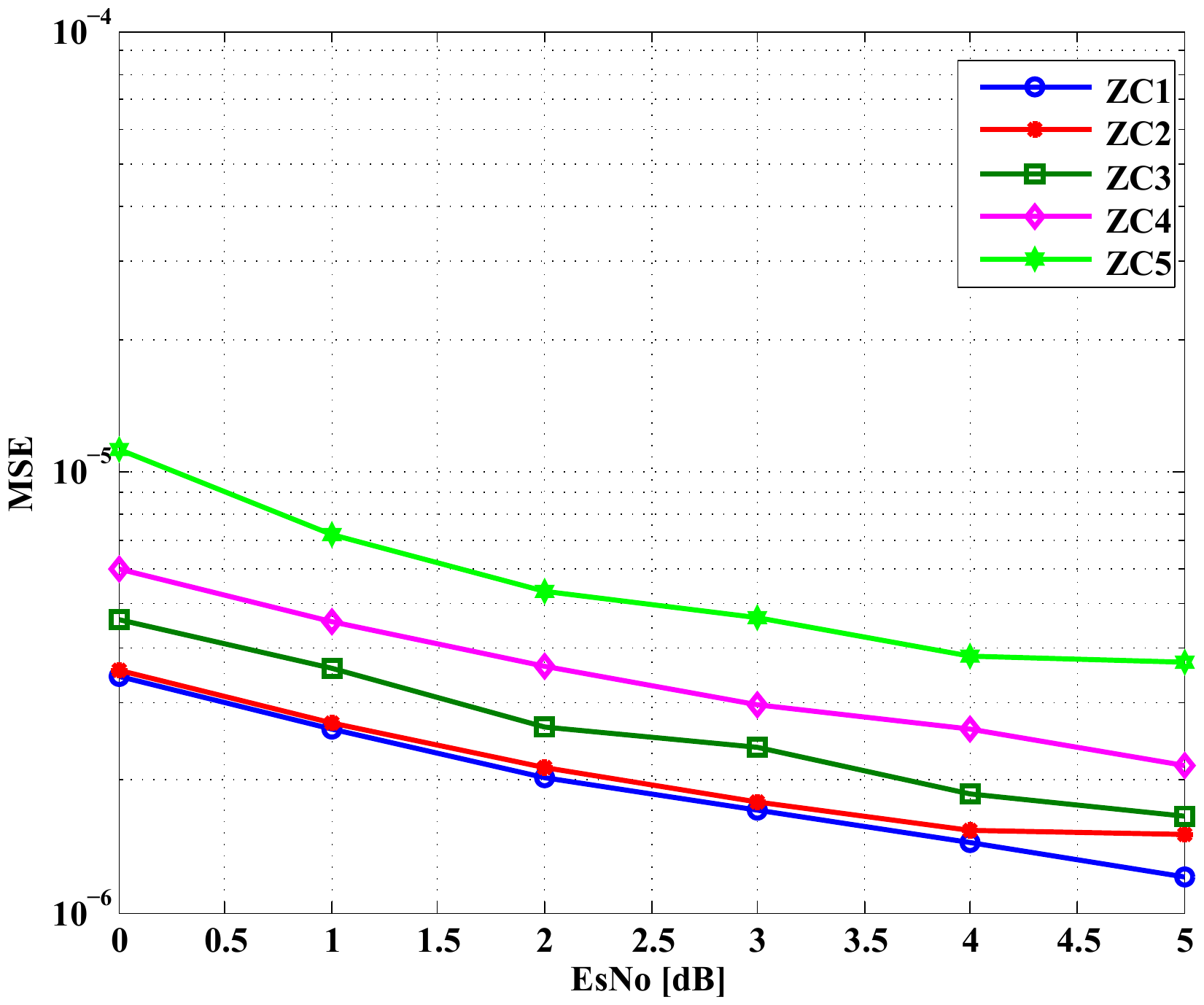}
\caption{MSE for estimated symbol rate, received data with observation period = 5e6, AWGN, Symbol Rate = 7 MSymbols/Sec}
\label{mse_5e6samples_LongRun_AWGN}
\end{figure}

In the case of received data with noise only, the farther zero crossings (4,5) will not be always the better ones as the noise effect on the farther zero crossing is larger than nearer one due to slope change. In figure \ref{mse_5e6samples_LongRun_AWGN}, we show the MSE for the SRE using first five zero crossings in case of received data with noise only. It is clear from the simulation in this case that the degradation due to noise is larger than the enhancements of dividing the interpolation error for the farther zero crossing (i.e., the noise error dominates at the observation length of 5e6 samples used in this case). So we have a performance trad-off between farther and nearer zero crossings dependent on the SNR region (including the noise averaging from the observation length). We should expect conversion to the performance limit in figure \ref{Without_Echo} while increasing the observation length. 

\begin{figure}
\centering
\includegraphics[width=3.2in]{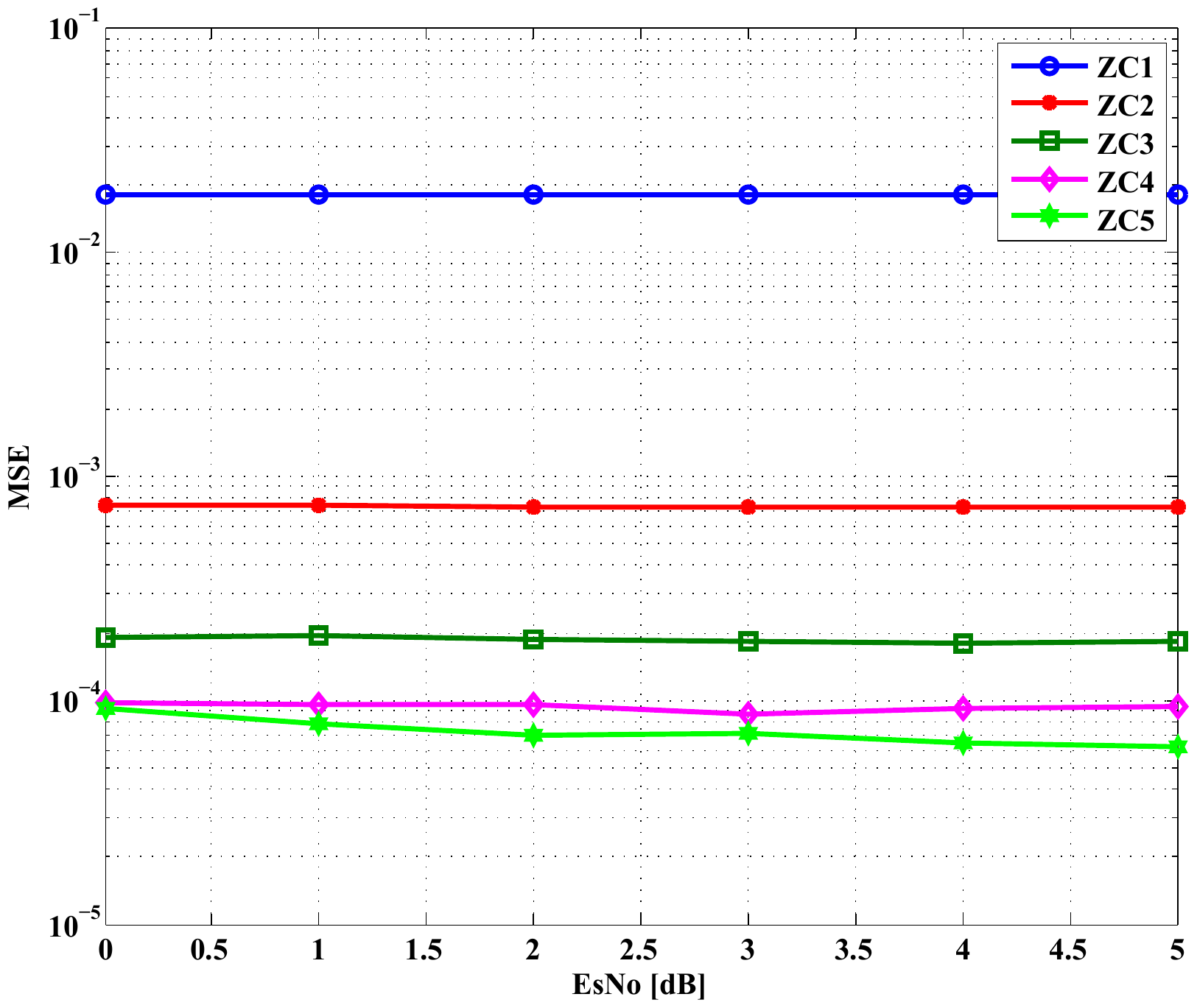}
\caption{MSE for estimated symbol rate, received data with observation period = 5e6, worst case echo channel, Symbol Rate = 7 MSymbols/Sec}
\label{mse_5e6samples_LongRun_Fading}
\end{figure}

In the case of practical received data with noise and channel, another factor appears here which is the error in estimation due to distortion from channel. In this case, the error due to the distortion dominates other errors, and hence, farther zero crossings always have the best performance since the dominant error due to channel distortion is significantly reduced. Figure \ref{mse_5e6samples_LongRun_Fading}, we show the MSE for the SRE using first five zero crossings in the case of received data with noise and worst case echo channel. In this case the farther zero crossing gives better MSE than earlier one (on the current SNR region including the observation length noise averaging). Note here that there is no trade-off here as the AWGN only case (or may be there is a trad-off but at very low SNR) as the behaviour for the SRE is the same as limit case in figure \ref{WorstCaseEcho_ZC_1to_7} (farther ZC is always better). Note that the MSE is about 1e-4 for ZC5 which is very close to the value in the limit case in figure \ref{WorstCaseEcho_ZC_1to_7} (1e4 ppm).

\subsection{Combining Different Zero Crossing Performance}

\begin{figure}
\centering
\includegraphics[width=3.2in]{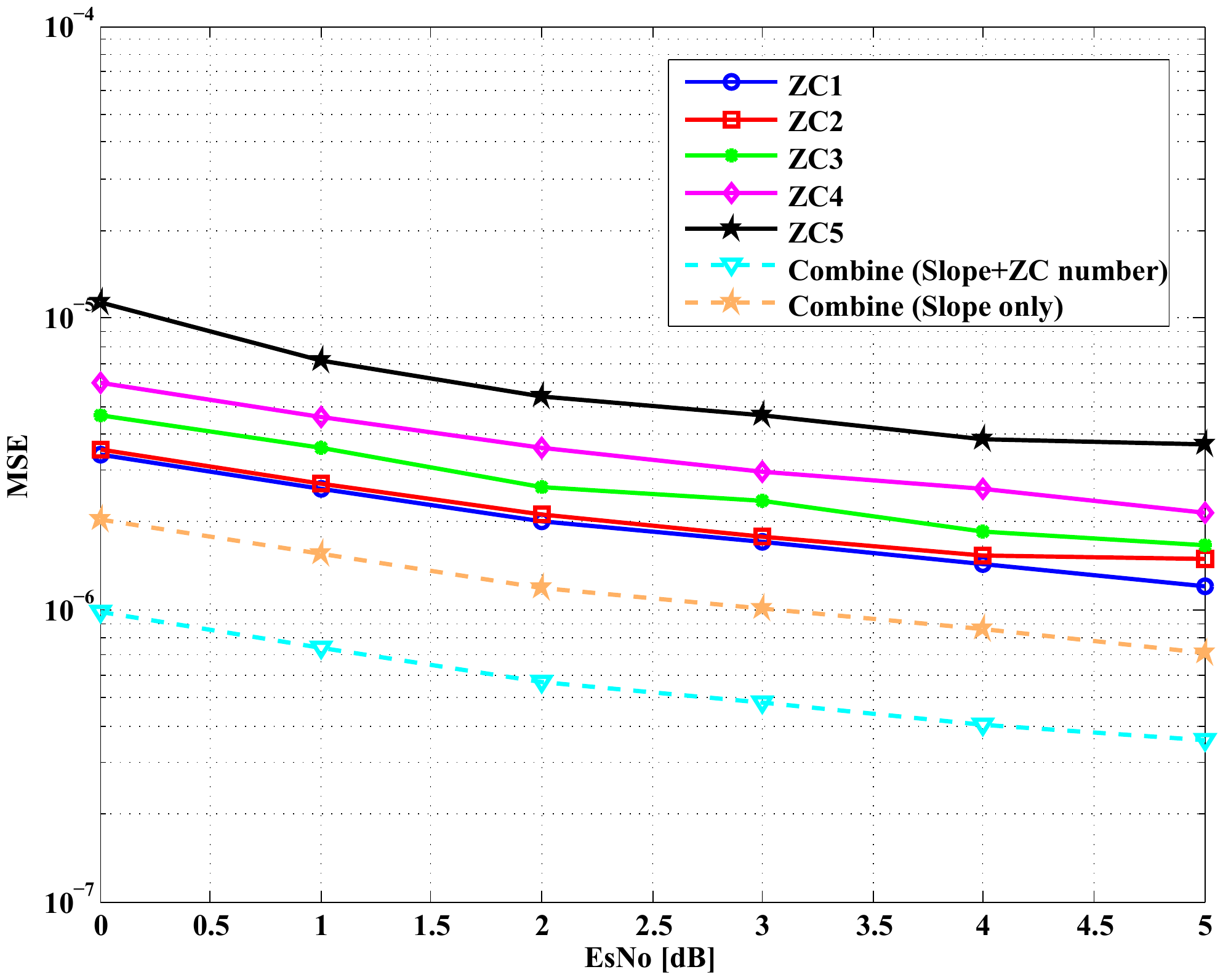}
\caption{MSE for estimated symbol rate after using combining, received data with observation period = 5e6, AWGN channel, Symbol Rate = 7 MSymbols/Sec}
\label{MSE_AWGN_LongLength_withCombine}
\end{figure}

Figure \ref{MSE_AWGN_LongLength_withCombine} shows the performance enhancements in the case of AWGN after combining with weights when using slope only and when using the slope plus zero crossing number. It is clear that weights that uses both slope and zero crossing number is better.

\begin{figure}
\centering
\includegraphics[width=3.2in]{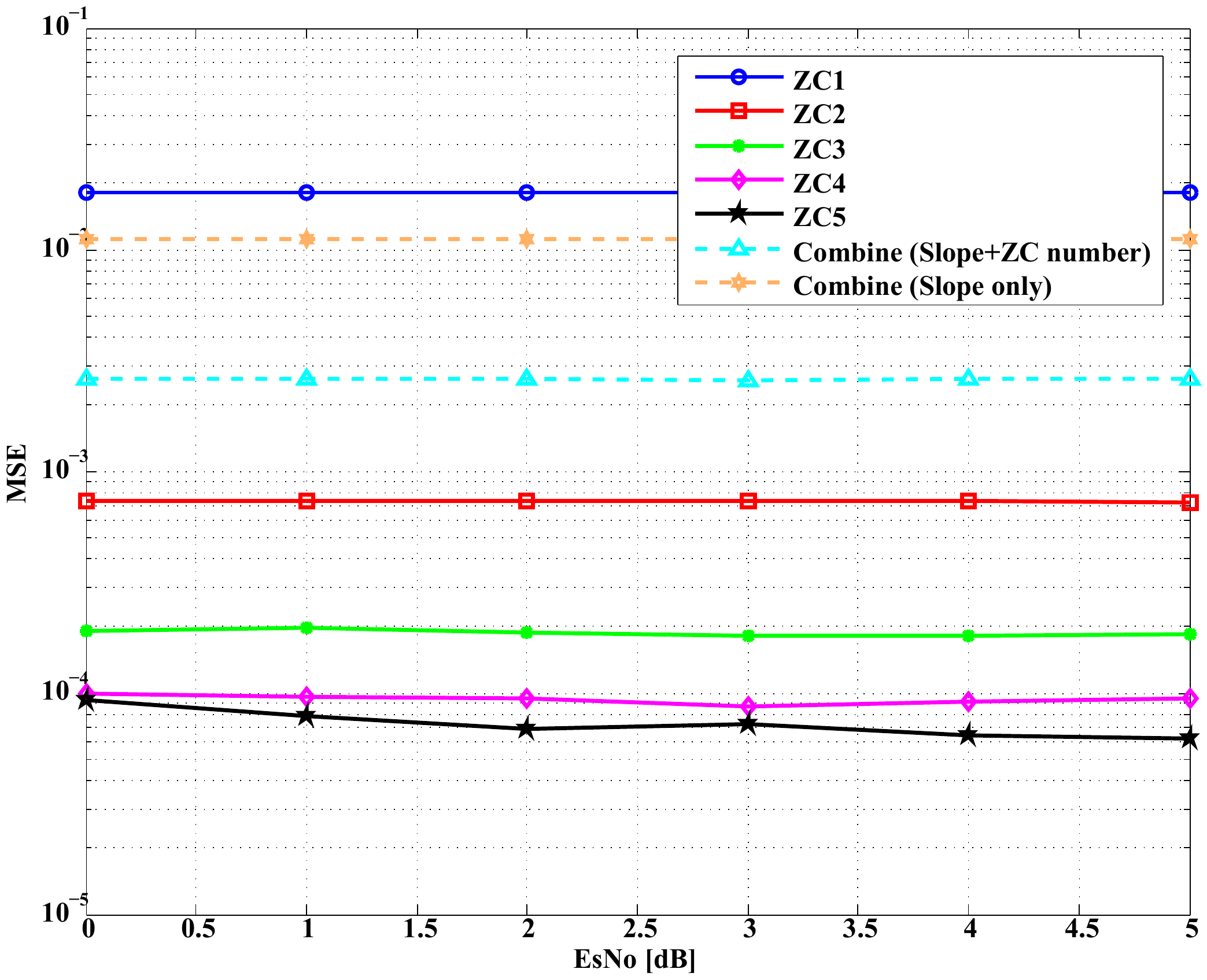}
\caption{MSE for estimated symbol rate after using combining, received data with observation period = 5e6, worst case echo channel, Symbol Rate = 7 MSymbols/Sec}
\label{MSE_Echo_LongLength_withCombine}
\end{figure}

The previous weights enhances the performance of the symbol rate estimator with a considerable amount in the case of AWGN. When testing the case of echo channel, we found that previous weights don't enhance the performance as the previous weights gives the nearer zero crossing the higher weight. However, in the channel case, the farther zero crossing should have the larger weight and this makes us think again on how to consider both slope and zero crossing number with proper relative scaling. The performance for the combining based on the previous weights for the echo channel is shown in figure \ref{MSE_Echo_LongLength_withCombine}.

\section{Conclusions and Discussion}

In this paper we presented a novel method for symbol rate estimation that depends on computing the time-averaged autocorrelation function and detecting the first zero crossing. The advantages of this method are its simplicity, good performance in low SNR situations, and the fact that it does not require any knowledge about modulation parameters. Furthermore, it is robust against low roll-off factors, which is a problem in many SRE methods in the literature. Additionally, the method does not detect a symbol rate from a candidate set, but produces an estimate from a continuum of values. The above gives the method wide applicability in the family of DVB systems as well as other applications. On the other hand, the method requires relatively long observation time, making it more immediately applicable to broadcast systems, where initial latency requirements are relatively relaxed. However, we believe this to be the case for many other SREs since they must perform to certain accuracy at the lowest SNR. Otherwise, the data communication stage will fail.

Improvements to the above method that may be investigated include least-squares curve fitting of the computed autocorrelation function with the RC pulse. This would require knowledge of the roll-off factor but would improve performance significantly. A variant may also be used where the roll-off factor is unknown or is known within a certain range or set, e.g. in DVB-S2 the candidate roll off factors are either 0.2, 0.25, or 0.35. Another important enhancement is to consider both the real and imaginary parts of the estimated Raised Cosine part as we use only here the real part.
Finally, the proposed algorithm has been implemented in RTL code and tested successfully on an FPGA using a third party tester.

\bibliographystyle{IEEEtran}

\bibliography{SRE_Journal}

\begin{thebibliography}{10}
\providecommand{\url}[1]{#1}
\csname url@samestyle\endcsname
\providecommand{\newblock}{\relax}
\providecommand{\bibinfo}[2]{#2}
\providecommand{\BIBentrySTDinterwordspacing}{\spaceskip=0pt\relax}
\providecommand{\BIBentryALTinterwordstretchfactor}{4}
\providecommand{\BIBentryALTinterwordspacing}{\spaceskip=\fontdimen2\font plus
\BIBentryALTinterwordstretchfactor\fontdimen3\font minus
  \fontdimen4\font\relax}
\providecommand{\BIBforeignlanguage}[2]{{%
\expandafter\ifx\csname l@#1\endcsname\relax
\typeout{** WARNING: IEEEtran.bst: No hyphenation pattern has been}%
\typeout{** loaded for the language `#1'. Using the pattern for}%
\typeout{** the default language instead.}%
\else
\language=\csname l@#1\endcsname
\fi
#2}}
\providecommand{\BIBdecl}{\relax}
\BIBdecl

\bibitem{ETSIDVBC}
EN300429V1.2.1, ``Digital video broadcasting (dvb); framing structure, channel
  coding and modulation for cable systems,'' April 1998.

\bibitem{Karam}
G.~Karam, K.~Maalej, V.~Paxal, and H.~Sari, ``Design and performance of a
  variable-rate qam modem for digital cable television,'' in \emph{Proc. of
  International Broadcasting Convention Conference}, Sept. 1995, p. 178 –
  183.

\bibitem{Lee}
J.~Lee, C.~Y., and S.~Lee, ``On a timing recovery technique for a variable
  symbol rate signal,'' in \emph{Proc. of Vehicular Technology
  Conference}.\hskip 1em plus 0.5em minus 0.4em\relax IEEE, May 1997, p.
  1724–1728.

\bibitem{Meyr}
H.~Wymeersch and M.~Moeneclaey, ``Ml-based blind symbol rate detection for
  multi-rate receivers,'' in \emph{Proc. of International Conference on
  Communications, ICC’05}.\hskip 1em plus 0.5em minus 0.4em\relax IEEE, May
  2005, p. 1724–1728.

\bibitem{Gardner}
W.~Gardner, ``Signal interception: A unifying theoretical framework for feature
  detection,'' \emph{Trans. On Communications}, vol.~36, pp. 897--906, August
  1988.

\bibitem{Dandawate}
A.~Dandawate and G.~Giannakis, ``Statistical tests for the presence of
  cyclostationarity,'' \emph{Trans. On Signal Processing}, vol.~42, pp.
  2355--2369, September 1994.

\bibitem{Mazet}
L.~Mazet and P.~Loubaton, ``Cyclic correlation based symbol rate estimation,''
  in \emph{Proc. of 33rd Asilomar Conf. on Signals, Systems, and Computers},
  vol.~2.\hskip 1em plus 0.5em minus 0.4em\relax IEEE, October 1999, pp.
  1008--1012.

\bibitem{Wanxue}
Y.~Wanxue and W.~Keren, ``A new method to symbol rate estimation of mpsk
  signals,'' in \emph{Proc. of Congress on Image and Signal}, vol.~5, May 2008,
  pp. 394--398.

\bibitem{Ciblat}
P.~Ciblat, P.~Loubaton, E.~Serpedin, and G.~Giannakis, ``Asymptotic analysis of
  blind cyclic correlation-based symbol-rate estimators,'' \emph{Trans. on
  Information Theory}, vol.~48, no.~7, p. 1922–1934, July 2002.

\bibitem{Mosquera}
C.~Mosquera, S.~Scalise, and R.~Valcarce, ``Non-data-aided symbol rate
  estimation of linearly modulated signals,'' \emph{Trans. On Signal
  Processing}, vol.~56, no.~2, p. 664 – 674, Febrauary 2008.

\bibitem{Chan}
Y.~Chan, B.~Lee, R.~Inkol, and F.~Chan, ``Estimation of symbol rate from the
  autocorrelation function,'' in \emph{Proc. of Canadian Conference on
  Electrical and Computer Engineering}, May 2009, pp. 547--550.

\bibitem{6655324}
M.~A. {Elgenedy} and A.~{Elezabi}, ``Blind symbol rate estimation using
  autocorrelation and zero crossing detection,'' in \emph{2013 IEEE
  International Conference on Communications (ICC)}, 2013, pp. 4750--4755.

\bibitem{Nordig}
NorDig, ``Nordig unified test specification, ver 2.2,'' 2010.

\bibitem{Deboor}
de~Boor, ``A practical guide to splines,'' 1978.

\end{thebibliography}

\end{document}